\documentclass[12pt,journal,letterpaper,onecolumn]{IEEEtran}

\usepackage{setspace,fleqn,url,epsf,calc,latexsym,amsmath,multicol,cite}
\usepackage[dvips]{graphics,color,graphicx}

\usepackage[ruled, vlined]{algorithm2e}

\newcounter{myCounter}
\renewcommand{\themyCounter}{\arabic{myCounter}\addtocounter{myCounter}{1}}

\newtheorem{theorem}{Theorem}[section]
\newtheorem{definition}{Definition}[section]
\newtheorem{example}{Example}[section]
\newenvironment{myproofsketch}
{\noindent {\bf Proof Sketch:}}{\hspace*{\fill}$\Box$}

\begin{document}

\title{Checking Security Policy Compliance}
 \begin{multicols}{2}
\author{Vaibhav Gowadia ~~~~~ Csilla Farkas  ~~~~~~~~~~~~~~Michiharu Kudo\\
       Center for Information Assurance Engineering~~~~~Tokyo Research Lab, IBM\\
     Department of Computer Science and Engineering~~~~~~~~~~~~~~~~~~~~~~~~~~~~~~~~~~~~\\
        University of South Carolina, Columbia ~~~~~~~~~~~~~~~~~~~~~~~~~~~~~~~~~~~~~\\
  }
  \end{multicols}
\date{January 30, 2008}

\maketitle 

\begin{abstract}
\noindent Ensuring compliance of organizations to federal
regulations is a growing concern. This paper presents a framework
and methods to verify whether an implemented low-level security policy is
compliant to a high-level security policy.  Our compliance checking framework
is based on organizational and security metadata to support
refinement of high-level concepts to implementation specific
instances. Our work uses the results of refinement calculus to
express valid refinement patterns and their properties.
Intuitively, a low-level security policy is compliant to a high-level security policy
if there is a valid refinement path from the high-level security policy to
the low-level security policy.  Our model is capable of detecting violations
of security policies, failures to meet obligations, and capability
and modal conflicts.
\end{abstract}


\section{Introduction}

\PARstart{R}{ecent} regulations, like Sarbanes-Oxley (SOX)~\cite{sox} and Health
Insurance Portability and Accountability Act (HIPAA)~\cite{Hipaa96},
are having a broad impact in information technology (IT) operations
at many organizations. For example, SOX requires organizations to
place adequate internal controls over financial reporting. HIPAA
requires sufficient safeguards to be placed for controlling access
to medical records. Moreover, these regulations require evaluations
of safeguards and controls implemented by the organizations to
determine whether they are compliant with the requirements.

Tools to support automated compliance checking and establish formal
properties are needed. Clearly, this is a complex problem requiring
knowledge not only about the high- and low-level policies but also
the available technologies, organizational requirements and
processes, and system dynamics. Several policy
languages~\cite{Damianou01,Kagal02,Jajodia01,bettini02} have been
proposed by researchers. However, they were not designed to allow
comparison of high-level and low-level security policies.

In this paper, we focus on the specific problem of checking
compliance of an implemented security policy to the high-level
security policy of an organization. A high-level security policy may
specify 1) description of security requirements over abstract
concepts, and 2) obligations, dispensations, and permissions. The
low-level security policy gives specific security requirements over
instances of abstract concepts. For example, let us consider an
organization with a business process called Order Management. A rule
in high-level policy may be that the Business Manager must protect
the Order Management process from unauthorized access. Rules in
low-level security policy may specify access control list for the
purchase orders database used by the Order Management process.

Refinement of a high-level policy into a low-level policy may
require instantiation of roles, refinement of actions, and inference
procedures.  Many researchers have also proposed mechanisms
for policy refinement~\cite{Bandara04,Rubio06,Rochaeli07},
i.e. to derive the low-level enforceable policies from the
high-level policies. Instantiation of roles has been studied extensively in
context of access control~\cite{Ferraiolo}.
The work of Backes et al.~\cite{Backes04} focuses on comparing two privacy policies.
However, the problem of verifying
compliance of a low-level implemented policy to a high-level policy
is not fully considered yet. In this paper, we propose a mechanism
based on refinement calculus~\cite{Back} to fill this gap.

In this work, we propose a policy refinement framework and
action algebra that we apply for checking compliance of security
policies. The proposed action algebra forms the basis of action
 refinements.  To illustrate the need
of action refinement to study compliance checking we now present an
example. Let $a_1$, $a_2$, and $a_3$ be actions, $s$ a subject, and
$o$ an object. Assume that allowing action $a_1$ is equivalent to
allowing action $a_2$ and disallowing action $a_3$. If a high-level
policy contains an access control rule $(s,o,+a_1)$ and low-level
policy contains access control rules $(s, o, +a_2)$ and $(s, o,
+a_3)$ then the low level policy is not compliant to the high-level policy.
Intuitively, the
policy compliance problem asks the question whether the low-level
policy satisfies the relevant requirements of the high-level
security policy.


Our main contributions in this paper are development of an action
algebra, a framework for policy refinement using refinement pattern,
and a definition of compliance based on the concept of model checking.
We describe a policy language that can model both high-level and low-level
security policies. The proposed policy language is an extension of
the Authorization Specification Language (ASL) and Flexible
Authorization Framework(FAF)~\cite{Jajodia01}. The
extended language supports specification of obligations,
dispensations, and authorizations. We have applied the principles of
refinement calculus to security policies, and developed an
action algebra that can be used to evaluate the correctness of
action compositions.  In addition, we have developed a policy
refinement mechanism that combines action algebra and the policy
language to refine high-level security policy into low-level
security policies. Security policies are refined using action
refinement patterns and derivation rules. The refinement process
results in a set of possible low-level policies and corresponding
system states. If the implemented low-level policy and the current
system state corresponds to a derived low-level policy and state
then we consider the implemented policy to be compliant to the
high-level policy.

Rest of the paper is organized as follows:
Section~\ref{sec:overview} presents an overview of the proposed
compliance checking framework. Section~\ref{sec:definitions}
presents definitions of basic constructs.
Section~\ref{sec:composition} describes action composition.
Section~\ref{sec:language} and \ref{sec:checking} describe our
extension of Flexible Authorization Framework(FAF) and the compliance checking process respectively.
In Section~\ref{sec:conclude} we conclude and recommend future work.


\section{Compliance Checking Framework}
\label{sec:overview}

\begin{figure}[h]
\begin{center}
$\begin{array}{c@{\hspace{.1in}}c}
\includegraphics[width=3in]{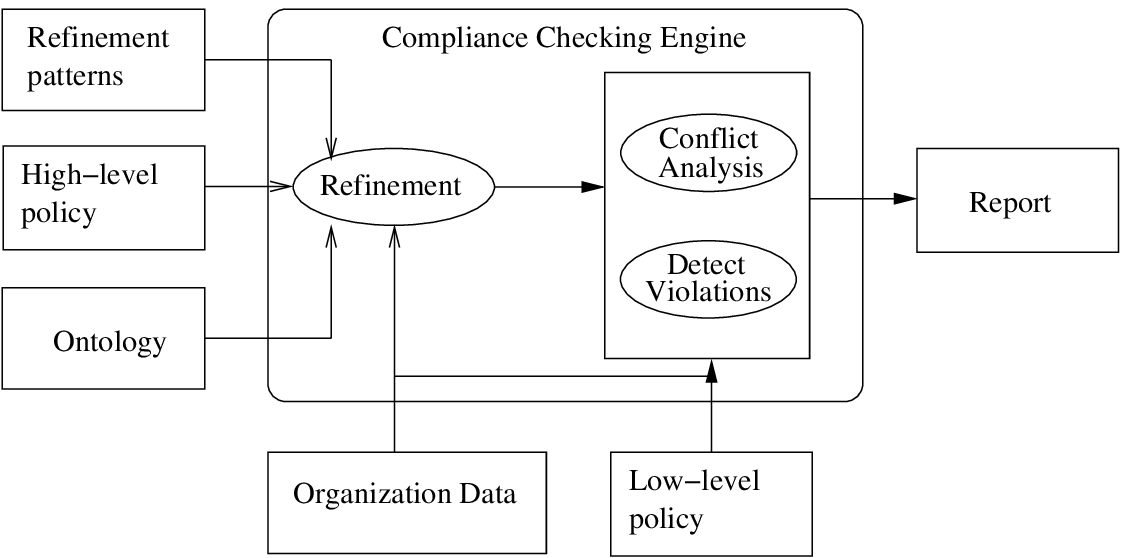} &
\includegraphics[height=2in]{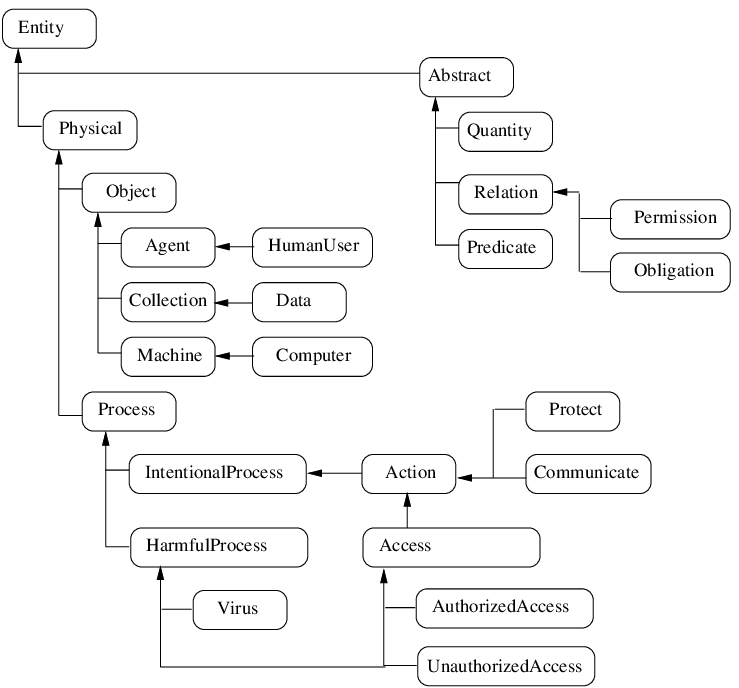}\\
 \mbox{\bf (a)} & \mbox{\bf (b)}
\end{array}$
\end{center}
 \caption{(a) Compliance checking architecture (b) Partial Ontology}
 \label{fig:overview}
\end{figure}

We propose a compliance-checking framework, where all entities in
the concerned organization are described with ontological concepts.
We define an ontology that models concepts like, subjects,
permissions, obligations, actions, protection objects, and metadata
associated with them and with the organization. Our compliance
checking framework comprises of the following components: 1. an
ontology, 2. instances of ontology concepts (e.g., users,
organization's resources, roles, etc.), 3. a high-level security
policy, 4. a set of low-level security policies, 5. refinement
patterns, and 6. compliance checking engine. An overview of the
compliance checking framework is shown in Figure~\ref{fig:overview}(a).
We now describe the components of the proposed compliance checking
framework.

We model security policies as locally stratified logic programs similar to
Authorization Specification Language~\cite{Jajodia01}. The
security policy language presented in this work can represent
obligation, dispensations, and authorizations. It also supports
conflict resolution rules and policy refinement. Action refinement
patterns specify refinement of an action of type $A$ into a
composition expression (Section~\ref{sec:composition}) formed with
sub-actions of $A$ such that the constraints for satisfying any
obligation of type $A$ are preserved.

The compliance checking engine in our framework refines the high-level security policy by recursively applying policy refinement rules. The refinement process continues until no new facts can be derived. The refined policies generated by this process comprise of ground rules and system-state information (facts) only. The set of all decision rules in a policy is called a decision view.

The low-level security policy and system information given as input to check for compliance is now compared with the set of refined security policies generated. If the given system state satisfies post conditions of applicable obligations and the decision view of input low-level policy implies one of the possible decision views of high-level policy, we say that the given system complies to high-level policy. However, if the given system is not compliant, the compliance checking engine may also detect violations of high-level policy and capability conflicts that prevent users from performing their obligations. In Section~\ref{sec:checking}, we discuss different types of violations and capability conflicts in further detail.

Rules in high-level policy contain composite actions. Composite
action consists of two or more sub-actions. We present an action
composition algebra and Ontology based system model to check whether
the action compositions are well-formed.

In next section, we define constructs used to model system state and
policy components like actions.

\section{Definitions}
\label{sec:definitions} This work uses ontologies to model the
entities of our compliance checking framework. Our method relies on this
ontology to aid the compliance checking as described in following
sections.  We now present our definition on Ontology used in
this work.


\begin{definition}(Ontology)\\
An ontology $O$ is a 6-tuple
$(\mathcal{C},\mathcal{P},\mathcal{C}_h,\mathcal{P}_h, dom, range)$,
where $\mathcal{C}$ is a set of classes, $\mathcal{P}$ is a set of
properties, $\mathcal{C}_h$ is the subclass hierarchy of
$\mathcal{C}$ and $\mathcal{P}_h$ is the subproperty hierarchy of
$\mathcal{P}$. $dom$ and $range$ are functions defined as
$dom:\mathcal{P}\rightarrow P(\mathcal{C})$ and
$range:\mathcal{P}\rightarrow P(\mathcal{C})$, where
$P(\mathcal{C})$ represents the power set of $\mathcal{C}$.
Let $c \in C$ be a class such that $c \in dom(p_i)$ for
$i=1,\ldots,k$ and let $r_i$ represent the $range(p_i)$ for
$i=1,\ldots,k$. We represent the class $c$ as
$c((p_1,r_1),\ldots,(p_k,r_k))$. \label{def:ontology}
\end{definition}

\begin{example} Let {\ttfamily Computer} be a class with properties {\ttfamily os}, {\ttfamily owner},
and {\ttfamily name}. Let the range of property {\ttfamily os} be
given by the class {\ttfamily OS}, the range of property {\ttfamily
owner} be given by the class {\ttfamily Agent}, and the range of
property {\ttfamily name} be given by the class {\ttfamily String}.
The class {\ttfamily Computer} is represented as {\ttfamily
Computer((os,\{OS\}),(owner,\{Agent\}),(name,\{String\}))}.
\label{eg:class}
\end{example}

Figure~\ref{fig:overview}(b) shows class hierarchy of the concepts
used in our framework. Our ontology is an extension of the SUMO
ontology~\cite{sumo} being developed by the IEEE SUO Working Group.
The root node of our ontology is the class {\em Entity}. The class
{\em Entity} refers to the fundamental concept in the domain being
modeled. The class {\em Object} refers to physical objects. Binary
relations that evaluates to true or false are represented by class
{\em Predicate}. {\em Process} is a class of active components that
occur and have temporal parts or stages. The class {\em Agent}
represents something or someone that can act on its own. For
example, software agents and human users. Human agents are
represented by the class {\em Users}. A set of users is called a
{\em Group}. A social position that is usually associated with some
obligations and permissions is called a {\em Role}. The class {\em
Action} represents a set of operations that the users may perform.
Properties of the class {\em Action} are shown in
Table~\ref{table:action}.

\begin{table}[h]
\begin{center}
\begin{minipage}{5in}
\footnotesize
\begin{tabular}{|l|l|p{3in}|}
\hline
Property & Range & Semantics \\
\hline
agent & Agent & Agent that actively carries out the process\\
instrument & Object & Instrument is used by the process and is not modified\\
resource & Object & Resource is modified and used by the process\\
target & Entity & The entity acted upon or modified by the process\\
evidence & Predicate & Predicate is true after the action is performed. \\
subAction & Action & A distinguished part of the process \\
causes & Process & This process causes or triggers another process of type specified by this property\\
prevents & Process &  Processes of type specified by this property are prevented by this process \\
\hline
\end{tabular}
\normalsize
\end{minipage}
\caption{OWL properties of class Action} \label{table:action}
\end{center}
\end{table}

\begin{definition}(Object)\\
Let $O$ be an ontology. An object is an instance of any class $c$
defined in $O$. Let $c((p_1,r_1),\ldots,(p_k,r_k))$ be the
definition of class $c$ where $p_1,\ldots, p_k$ are properties
of class $c$. We describe object $o$ as
$o((type,c),$ $(\overline{p_1},v_1),\ldots,(\overline{p_l},v_l))$,
where $o$ is a unique identifier, and for each $\overline{p_i}$
there is a $p_j$ of $c$ such that $\overline{p_i}=p_j$
and $v_i$ is in range $r_j$. We use the notation
$p_i(o,v_i)$ to represent the $i^{th}$ property of $o$ and
its value. Note that {\ttfamily type} is one of the properites
of $o$. \label{def:object}
\end{definition}

\begin{example} An object of type {\ttfamily Computer} with
os {\ttfamily Solaris8}, owner {\ttfamily Alice}, and name
{\ttfamily Hadar} is represented as {\ttfamily
Comp1((type,Computer),(os,Solaris8)},{\ttfamily(owner,Alice)}, {\ttfamily(name,Hadar))},
where {\ttfamily Comp1} is an identifier used to represent the
computer object in question. Also note that {\ttfamily Solaris8},
{\ttfamily Alice}, and {\ttfamily Hadar} are identifiers of other
objects. \label{eg:object}
\end{example}

\begin{definition}(Data System)\\
The Data System $DS$ = \{$o_1,\ldots,o_n$\} is a set of objects.
  \label{def:ds}
\end{definition}

\begin{definition}(State)\\
The state of a data system $DS$ is described by properties of
objects in $DS$, that is
$\{p_1^1(o_1,v_1),\ldots,$ $p_k^1(o_1,v_k)\}$ $\cup \ldots \cup$
$\{p_1^n(o_n,v^n_1),\ldots,p_m^n(o_n,v^n_m)\}$.
 \label{def:state}
\end{definition}

For simplicity, in
the rest of this paper, we represent $p_i(o_j,v_i)$ as $x_i=v_i$,
where $x_i$ is a variable representing the property $p_i$ of object
$o_j$. We say that the range of $x_i$ is the same as the range of
$p_i$. Let $X = (x_1,\ldots,x_i,\ldots,x_{h})$ be the set of
variables that describe a state in $DS$. The mapping from  $X$ to
objects and their properties is maintained separately.

 Alternatively, a state $\gamma$ is defined as an assignment
$x_1:=v_1, \ldots, x_h:=v_h$, where $v_i~(i=1,\ldots,h)$ is value of
variable $x_i$ and $v_i \in r_i$.

Note that a system may satisfy more than one state representations.
These state representations are related to each other by refinement
relation as we describe below.
\begin{definition} (State Refinement)\\
\noindent Let $\gamma = \{x_1:=v_1, \ldots, x_n:=v_n\}$ and $\gamma' =
\{x_1:=v'_1, \ldots, x_n:=v'_n\}$ be two states. We say that $\gamma'$ is
a refinement of $\gamma$ ($\gamma \sqsubseteq \gamma'$) if $v'_1 \leq_h v_1, \ldots,
v'_n \leq_h v_n$. Note that the refinement relation ($\sqsubseteq$)
between states is reflexive, transitive, and antisymmetric.
\end{definition}

\begin{example} Let  $\gamma$ = \{$x_1$:={\ttfamily Computer},
$x_2$:={\ttfamily Linux\}}, $\gamma'$= \{$x_1$:={\ttfamily Notebook}, $x_2$:={\ttfamily Linux\}} be two state representations for an object. Given,
{\ttfamily Notebook} $\leq_h$ {\ttfamily Computer}, we can say that
state $\gamma$ is refined by state $\gamma'$ $(\gamma \sqsubseteq \gamma')$. \label{eg:state}
\end{example}

\begin{definition}(State Space)\\
A state space is a set of states.
 \label{def:statespace}
\end{definition}

\begin{example} Let $\gamma_1$ = \{$x_1$:={\ttfamily Computer}, $x_2$:={\ttfamily Linux}\},
     $\gamma_2$ = \{$x_1$:={\ttfamily Computer}, $x_2$:={\ttfamily Windows}\} be two states.
     Then the set $\Gamma = $\{$\gamma_1, \gamma_2$\} represents a state space.
\label{eg:statespace1}
\end{example}

Description of a state space as illustrated
in above example can be very tedious for large systems. In many cases,
we want to specify only the variables of interest. We allow a more concise
description of a state space in such cases as described below.

Let $DS$ be a data system that can be described by variables
$x_1, \ldots, x_n$, where $range(x_i)=r_i$ $(i=1,\ldots,n)$.
A state space $\Gamma$  described as $(x_1=v_1,
\ldots, x_k=v_k)~(k \leq n)$ represents the following set of states:\\
$\{x_1:=v_1\}$ x $\{x_2:=v_2\}$ x $\ldots \{x_k:=v_k\}$ x
$\{x_{k+1}:=v_{k+1}^1,
 x_{k+1}:=v_{k+1}^2,\ldots,x_{k+1}:=v_{k+1}^m \}$ x
   $\ldots$ x
 $\{x_n:=v_n^1,x_n:=v_n^2,\ldots,x_n:=v_n^l\}$
 where,  $range(x_{k+1}) = \{v_{k+1}^1,\ldots, v_{k+1}^m\}$, $\ldots$ ,
 $range(x_n)=\{v_n^1,\ldots,v_n^l\}.$

\begin{example}
Let us assume data system $DS$ contains only two variables $x_1$ and $x_2$.
Let state space $\Gamma$ be described as ($x_1$:={\ttfamily Computer}), and
 $range(x_2)$=\{{\ttfamily Linux, Windows}\}. \\
 Then $\Gamma$ = \{$x_1$:={\ttfamily Computer}\} x \{$x_2$:={\ttfamily Linux}, $x_2$:={\ttfamily Windows}\}, \\
 i.e., $\Gamma$ = \{ \{$x_1$:={\ttfamily Computer}, $x_2$:={\ttfamily Linux} \},
                     \{$x_1$:={\ttfamily Computer}, $x_2$:={\ttfamily Windows}\}
                   \}
\label{eg:statespace2}
\end{example}

Intuitively, refinement of a state space means reaching a more
specific state space. A more specific state space has fewer states or
contains states that are sub states of states in other state space.
Refinement of state space is now formally defined.

\begin{definition}(State Space Refinement)\\
Let $\Gamma$ and $\Gamma'$ be two state spaces. We say that $\Gamma$
is refined by $\Gamma'$ ($\Gamma \sqsubseteq \Gamma'$), if and only
if $\forall \gamma' \in \Gamma'$,  $\exists \gamma \in \Gamma$
such that $\gamma \sqsubseteq \gamma'$.
\label{def:statespacerefinement}
\end{definition}

\begin{example}
Let us assume $\gamma_1$ = \{$x_1$:={\ttfamily Computer}, $x_2$:={\ttfamily Linux}\},
     $\gamma_2$ = \{$x_1$:={\ttfamily Computer}, $x_2$:={\ttfamily Windows}\}, and
     $\gamma_3$ = \{$x_1$:={\ttfamily Notebook}, $x_2$:={\ttfamily Linux}\} are states, and $\Gamma_1$ = \{$\gamma_1$\},
     $\Gamma_2$ = \{$\gamma_1, \gamma_2$\}, and
     $\Gamma_3$ = \{$\gamma_3$\} describe state spaces.
From definition of state space refinement, we observe
1) $\Gamma_2$ is refined by $\Gamma_1$ ($\Gamma_2 \sqsubseteq \Gamma_1$), as $\gamma_1 \in \Gamma_1$, $\gamma_1 \in \Gamma_2$, and $\gamma_1 \sqsubseteq \gamma_1$ and 2) $\Gamma_1$ is refined by $\Gamma_3$ ($\Gamma_1 \sqsubseteq \Gamma_3$), as $\gamma_3 \in \Gamma_3$, $\gamma_1 \in \Gamma_1$ and $\gamma_1 \sqsubseteq \gamma_3$.
\label{eg:statespaceref}
\end{example}

The refinement relation between states spaces is reflexive,
transitive, and antisymmetric.
\begin{center}
\begin{minipage}{4.5in}
$\Gamma \sqsubseteq \Gamma$  (reflexive) \\
$\Gamma \sqsubseteq \Gamma' $ $~\&~$ $ \Gamma' \sqsubseteq \Gamma''
~\Rightarrow~ \Gamma \sqsubseteq \Gamma''$ (transitive)\\
$\Gamma \sqsubseteq \Gamma' $ $~\&~$ $ \Gamma' \sqsubseteq \Gamma
~\Rightarrow~ \Gamma = \Gamma'$ (antisymmetric)
\end{minipage}
\end{center}

Let $\Sigma$ represent a non empty state space that contains all
possible states of data system $DS$, and $P(\Sigma)$ be the power set
of $\Sigma$. The pair ($P(\Sigma),\sqsubseteq$) is then a partially
ordered set. Let $\Gamma$ and $\Gamma'$ be two elements (state
spaces) in $P(\Sigma)$. The greatest lower bound of $\Gamma$ and
$\Gamma'$ is given as $\Gamma \sqcap \Gamma' = \Gamma \cap \Gamma'$.
The least upper bound of $\Gamma$ and $\Gamma'$ is given as $\Gamma
\sqcup \Gamma' = \Gamma \cup \Gamma'$.

\begin{definition}(Restricted Subclass)\\
Let $c((p_1,r_1),\ldots,(p_n,r_n))$ be a class. Then
$c((p^i_1,r'_1),\ldots,(p^i_j,r'_j))$ is a restricted subclass,
where at least one of $r'_1,\ldots$, $r'_j$ is a subclass or an
instance of $r_1,\ldots,r_k$ respectively. For all other $r'_i$,
$r'_i=r_i$. \label{def:RestrictedSubclass}
\end{definition}

\begin{example}  Consider the class {\ttfamily Computer} defined
in Example~\ref{eg:class}. The restricted class {\ttfamily
Computer} {\ttfamily((os,Windows))} represents the sub class comprising of all
computers with operating system of type {\ttfamily Windows}.
\end{example}

\begin{definition}(Action)\\
\label{def:action} Let $\Delta$ and $\Gamma$ be two state spaces. An
action class $A: \Delta \rightarrow \Gamma$ is a state transformer
from $\Delta$ to $\Gamma$. An action $a:\delta \rightarrow \gamma$
is an instance of action class $A$ only if $\delta \in \Delta$ and
$\gamma \in \Gamma$. We call $\Delta$ as the initial state space and
$\Gamma$ as the final state space for action class $A$.
\end{definition}

In the rest of this paper, for each variable $A_i$ of type action class,
 we assume there exists corresponding initial and final state spaces
 and we use symbols $\Delta_i$ and $\Gamma_i$ to denote them.

\begin{definition}(Monotonicity of Refinement)\\
\label{def:monotonicity}
Let $a:\Delta \rightarrow \Gamma$ be an action. Let \{$\delta$\} and
\{$\delta'$\} be state spaces such that
$\Delta \sqsubseteq \{\delta\} \sqsubseteq \{\delta'\}$. If
$a(\delta) \rightarrow \gamma$ and $a(\delta') \rightarrow \gamma'$,
then $\gamma \sqsubseteq \gamma'$.\\

Let $a:\Delta\rightarrow\Gamma$ be an action.
Let $\Delta'=\{\delta'_1,\ldots,\delta'_n\}$ be a state space that
refines $\Delta$ ($\Delta \sqsubseteq \Delta'$).
Let $a(\delta'_i) \rightarrow \gamma'_i$ represent actions performed
 on states in the state space $\Delta'$, and let
 $\Gamma' = \{\gamma'_1, \ldots, \gamma'_n\}$ be the state space
 after action $a$ is performed.  By definition of $A$, the state
 space $\Gamma$ is refined by $\gamma'_i$ $(i=1,\ldots,n)$. This
 implies that $\Gamma \sqsubseteq \Gamma'$.

\end{definition}

For better readability we often write
$\Gamma \sqsubseteq \gamma$ instead of $\Gamma \sqsubseteq \{\gamma\}$
in rest of the paper.

Actions are often composed of several other sub-actions. Composition
may be performed by following operations: sequence ($;$), choice
($\vee$), and conjunction ($\wedge$). These operators give us the
following language for
expressing action composition: \\
\indent $A := a ~|~ (A) ~|~ A_1 ; A_2 ~|~ A_1 \vee A_2 ~|~ A_1 \wedge A_2$\\
where $a$ is an atomic action, and $A_1$ and $A_2$ are subactions of
$A$.  The precedence order of the operators in action composition is
$() > \wedge > ; > \vee$. We now describe properties of these
operators.

The choice operator ($\vee$) is a binary operator. If $a_1$ and
$a_2$ represent two action terms then $a_1 \vee a_2$ represents an
 action $a$ that executes either $a_1$ or
$a_2$. The choice operation is commutative and associative.
\begin{center}
\begin{minipage}{4.5in}
\begin{multicols}{2}
$a_1 \vee a_1 \equiv a_1$ \\
$a_1 \vee a_2 \equiv a_2 \vee a_1$ \\
$(a_1 \vee a_2) \vee a_2 \equiv a_1 \vee (a_2 \vee a_3)$
\end{multicols}
\end{minipage}
\end{center}

The sequence operator ($;$) is also a binary operator. If $a_1$ and
$a_2$ represent two action terms then $a_1;a_2$ represents an action
that performs $a_1$ followed by $a_2$. The sequence operator is not
commutative. It is associative and is distributive over the choice
operator.
\begin{center}
\begin{minipage}{4.5in}
\begin{multicols}{2}
$a_1;\{\} \equiv \{\};a_1 \equiv a_1$\\
$a_1 ; a_2 \neq a_2 ; a_1$ \\
$(a_1;a_2);a_3 \equiv a_1; (a_2;a_3)$ \\
$(a_1 \vee a_2);a_3 \equiv a_1;a_3 \vee a_2;a_3$\\
$ a_3;(a_1 \vee a_2) \equiv a_3;a_1 \vee a_3;a_2$
\end{multicols}
\end{minipage}
\end{center}

Choice and sequence operators are the basic operators in our action
algebra. The conjunction operator is a composite operator. If $a_1$
and $a_2$ are two action terms, then a conjunction operation on
$a_1$ and $a_2$ is defined by $a_1 \wedge a_2 = a_1;a_2 \vee
a_2;a_1$. Conjunction operator is associative, commutative
 and is distributive over the choice operator.
\begin{center}
\begin{minipage}{4.5in}
\begin{multicols}{2}
$a_1 \wedge a_2 \equiv a_2 \wedge a_1$ \\
$a_1 \wedge (a_1 \vee a_2) \equiv a_1 \wedge a_2 \vee a_1 \wedge
a_3$\\
$(a_1 \wedge a_2) \wedge a_3 \equiv a_1 \wedge (a_2 \wedge a_3)$
\end{multicols}
\end{minipage}
\end{center}



\begin{definition}(Action Refinement)\\
Let $a:\Delta \rightarrow \Gamma$, $a_1:\Delta_1 \rightarrow
\Gamma_1$ and $a_2:\Delta_2 \rightarrow \Gamma_2$ be actions, and
 $a_1 \oplus a_2$ be an action composition. We say that $a_1
\oplus a_2$ is a refinement of $a$, i.e., $a \sqsubseteq a_1 \oplus
a_2$, iff given any states $\delta \in \Delta$ and $\gamma \in
\Gamma$, where $a(\delta) \rightarrow \gamma$, the action
composition $(a_1\oplus a_2)(\delta) \rightarrow \gamma'$, such that
$\gamma \sqsubseteq \gamma'$. \label{def:action-refinement}
\end{definition}

We assume that the action composition $a_1\oplus a_2$ used
as a pattern to refine an action $a$ in our model is alway more 
restrictive then $a$. Therefore, it not possible for the refinement
process to derive $a$ when further refining the composition 
$a_1\oplus a_2$.

\begin{definition}(Atomic Action)\\
 An action $a$ is an atomic action if it
cannot be refined by any other action.
\end{definition}

\begin{definition}(Action Tree)\\
An action composition tree is a node-labeled binary tree where each
internal node is labeled with an action and an operator pair, and
each leaf node is labeled with an atomic action. The composition of
actions represented by the child nodes is a refinement of the action
at the parent node.
\end{definition}

In next section, we present types of action compositions that are
allowed in this work. To provide assurance about correct compliance
checking and policy refinement, action compositions must be
well-formed. The concept of well-formed action composition is also
discussed in next section.

\section{Action Composition}
\label{sec:composition} We first define types of action compositions
categorized based on the depth of action tree.

\begin{definition}(Simple Composition)\\
Let $a_1$ and $a_2$ be two atomic actions and $\oplus$ be an action
composition operator. An action composition of the form $a_1 \oplus
a_2$ is called a simple composition. \label{def:simple}
\end{definition}

\begin{definition}(Complex Composition)\\
Let $a_1$ and $a_2$ be two actions and $\oplus$ be an action
composition operator. An action composition $a_1 \oplus a_2$ is a
complex composition if 1.) $a_1$ and $a_2$ are either
 atomic actions, simple compositions, or complex compositions, and
 2.) at least one of $a_1$ and $a_2$ is not an atomic action.
\end{definition}

An action refinement pattern is a template for refining actions of a
particular type.

\begin{definition}(Action Refinement Pattern)\\
A refinement pattern $\mathcal{RP(A)}$ is an action tree with root
node of action type $\mathcal(A)$.
\end{definition}

We define additional types of composition in the context of
refinement. We categorize action compositions as basic or
advanced based on absence or presence of constraints in addition to
operator type. Note that the advanced composition type is applicable
only when it is required to perform both sub-action. Hence, it is
not applicable to choice operations.

\begin{definition}(Basic Composition)\\
\label{def:basic} Let $a_1$ and $a_2$ be two actions. We say that
$a\sqsubseteq a_1 \oplus a_2$ is a basic composition if $\oplus$ is
one of the
 operators: sequence, choice, or conjunction as defined in
Section~\ref{sec:definitions} and there are no additional
constraints.
\end{definition}

\begin{definition}(Advanced Composition)\\
\label{def:advanced}  Let $a_1$ and $a_2$ be two actions, $\oplus$
be an operator, and $\Delta'$ be a state space such that $\Delta_2
\sqsubseteq \Delta'$ (or $\Delta_1 \sqsubseteq \Delta'$). We say
that $a\sqsubseteq a_1 \oplus [\Delta']a_2$ (or $a\sqsubseteq
[\Delta']a_1 \oplus a_2$) is an advanced composition if for all
states in $\Delta'$ the sub action $a_2$ (or $a_1$ respectively) can
be ignored but for all states in $\Delta - \Delta'$ both $a_1$ and
$a_2$ must be performed.
\end{definition}

We also categorize action compositions as strict or flexible
based on the feasibility to perform both sub-actions in the initial
state space or the feasibility to perform at least one of
sub-actions in the initial state space. Strict and flexible action
composition types are not applicable for sequence operators as the
order of sub-actions is predetermined.

\begin{definition}(Strict Composition)\\
\label{def:strict} Let $a \sqsubseteq a_1 \oplus_s a_2$, where
$\oplus_s$ represents a composition operator. We say that $a_1
\oplus_s a_2$ is a strict composition if $\Delta$ is constrained
strictly to satisfy conditions such that {\em both} $a_1$ and $a_2$
can be performed in the initial state for all $\delta \in \Delta$.
In other words, $\Delta_1 \sqcap \Delta_2 \sqsubseteq \Delta$.
\end{definition}

\begin{definition}(Flexible Composition)\\
\label{def:flexible} Let $a \sqsubseteq a_1 \oplus a_2$, where
$\oplus$ represents a composition operator. We say that $a_1 \oplus
a_2$ is a flexible composition if for all $\delta \in \Delta$, it is
feasible to perform {\em either} $a_1$ or $a_2$ in the initial
state, and both $a_1$ and $a_2$ must be performed. i.e., $\Delta_1
\sqcup \Delta_2 \sqsubseteq \Delta$.
\end{definition}

The above composition types may be combined.
Figure~\ref{fig:compositionTypes} illustrates possible combinations.
Constraints for correct action refinements for various composition
types are given in Table~\ref{table:refine}.

\begin{figure}
\begin{minipage}{6in}
\footnotesize
\begin{center}
\begin{tabular}{|l|l|l|}
\hline
& Basic & Advanced \\
\hline
Sequence & $a \sqsubseteq a_1;a_2$ & $a \sqsubseteq a_1;[\Delta']a_2$ \\
\hline
 Strict & $a \sqsubseteq a_1 \wedge_s a_2$ & $a \sqsubseteq
a_1
\wedge_s [\Delta'] a_2 $ \\
 & $a \sqsubseteq a_1 \vee_s a_2$ & \\
\hline
Flexible & $a \sqsubseteq a_1 \wedge a_2$ & $a\sqsubseteq a_1 \wedge [\Delta']a_2$ \\
& $a\sqsubseteq a_1 \vee a_2$ & \\
\hline
\end{tabular}
\end{center}
\normalsize
\end{minipage}
\caption{Composition Types} \label{fig:compositionTypes}

\end{figure}

\begin{table}[h]
\begin{center}
\begin{minipage}{6in}
\footnotesize
\begin{center}
\begin{tabular}{|p{3.2cm}|llll|}
\hline
 Composition Type &             &&&\\
\hline \hline
Basic &   Constraints          &&&\\
\hline

$a \sqsubseteq a_1 ; a_2$ &

$\Delta_1 \sqsubseteq \Delta$ &

$\Delta_2 \sqsubseteq \Gamma_1$ &

$\Gamma \sqsubseteq \Gamma_2$ & \\
\hline \hline
 Basic and Strict & Constraints &&&\\
\hline &
\multicolumn{3}{l}{Let $\delta \in \Delta$ be the start state.} & \\

$a \sqsubseteq a_1 \vee_s a_2 $ &

$ \Delta_1 \sqcap \Delta_2 \sqsubseteq \Delta$ &

 $ \Gamma \sqsubseteq \Gamma_1$ &

$\Gamma \sqsubseteq \Gamma_2$ & \\

$a \sqsubseteq a_1 \wedge_s a_2$ &

$\Delta_1 \sqsubseteq \Delta$ & $\Delta_2 \sqsubseteq \Delta$ && \\

& $\Gamma  \sqsubseteq a_1(a_2(\delta))$ & $\Gamma \sqsubseteq a_2(a_1(\delta))$ && \\

\hline \hline
Basic and Flexible & Constraints &&& \\
\hline

$a \sqsubseteq a_1 \vee a_2 $ &

$ \Delta_1 \sqcup \Delta_2 \sqsubseteq \Delta$ &

$\Delta_1 \sqcap \Delta \neq \{ \} $ &

$\Delta_2 \sqcap \Delta \neq \{ \}$ &\\

& & $ \Gamma \sqsubseteq \Gamma_1 $& $ \Gamma
\sqsubseteq \Gamma_2$ & \\

$a \sqsubseteq a_1 \wedge a_2$ &

$ \Delta_1 \sqcup \Delta_2 \sqsubseteq \Delta$ &

\multicolumn{2}{l}{$ \Delta_1 \sqsubseteq \delta \Rightarrow
\Delta_2 \sqsubseteq
a_1(\delta)$ and $\Gamma \sqsubseteq a_2(a_1(\delta))$}& \\

&& \multicolumn{2}{l}{$ \Delta_2 \sqsubseteq \delta \Rightarrow
\Delta_1 \sqsubseteq
a_2(\delta)$ and $\Gamma \sqsubseteq a_1(a_2(\delta))$} &\\

\hline
\hline Advanced &   Constraints          &&&\\
\hline

$ a \sqsubseteq a_1 ; [\Delta'] a_2$ & $ \Delta_1 \sqsubseteq \Delta
$  &

$ \Delta' \not \sqsubseteq a_1(\delta) \Rightarrow \Gamma
\sqsubseteq a_1(\delta)$ & & \\

&& \multicolumn{2}{l}{$\Delta' \sqsubseteq a_1(\delta) \Rightarrow \Gamma \sqsubseteq a_2(a_1(\delta))$}&\\

\hline \hline
Advanced and Strict & Constraints &&& \\
\hline

$ a \sqsubseteq a_1 \wedge_s [\Delta'] a_2$ &

$\Delta_1 \sqsubseteq \Delta$ & $\Delta \sqcap \Delta' \neq \{\}$
&$\Delta \sqcap \Delta' \not \sqsubseteq \delta \Rightarrow$
 $\Gamma \sqsubseteq a_1(\delta)$&\\

& \multicolumn{3}{l}{$\Delta \sqcap \Delta' \sqsubseteq \delta
\Rightarrow$
  $\Delta_1 \sqsubseteq \delta$ and $\Delta' \sqsubseteq
  a_1(\delta)$ and $\Gamma \sqsubseteq a_2(a_1(\delta))$} &\\

& \multicolumn{3}{l}{$\Delta \sqcap \Delta' \sqsubseteq \delta
\Rightarrow$
 $\Delta_1 \sqsubseteq a_2(\delta)$ and $\Gamma \sqsubseteq
 a_1(a_2(\delta))$} &\\

\hline
\hline Advanced and Flexible & Constraints &&& \\
\hline

$ a \sqsubseteq a_1 \wedge [\Delta'] a_2$ &


$\Delta_1 \sqcup \Delta' \sqsubseteq  \Delta$ &  && \\

 &\multicolumn{3}{l}{$\Delta_1 \sqsubseteq \delta$ and
$ \Delta' \sqsubseteq a_1(\delta)$ $\Rightarrow$ $\Gamma \sqsubseteq a_2(a_1(\delta))$} & \\

 &\multicolumn{3}{l}{$ \Delta_1 \sqsubseteq \delta$ and $\Delta'
\not \sqsubseteq \delta$ and $\Delta' \not \sqsubseteq a_1(\delta)
\Rightarrow \Gamma
\sqsubseteq a_1(\delta)$} & \\

&\multicolumn{3}{l}{$\Delta' \sqsubseteq \delta$ and $\Delta_1
\sqsubseteq a_2(\delta)$ $\Rightarrow \Gamma \sqsubseteq
a_1(a_2(\delta))$} & \\

 \hline
\end{tabular}
\end{center}
\normalsize
\end{minipage}
\caption{Constraints for well-formed action refinement}
\label{table:refine}
\end{center}
\end{table}

\begin{definition}(Valid Action Trace)\\
Action trace is an action composition where the composition is
expressed using only the sequence operator. Given an action tree and
an action trace, the trace is valid with respect to the action tree
iff the trace can be derived from the root of the action tree using
the properties of the operators.
\end{definition}

\begin{definition}(Well-formed Action Composition)\\
Let $a \sqsubseteq a_1 \oplus a_2$ be an action composition and let
$T_1, \ldots, T_n$ represent all the valid traces of the action
composition. We say that an action composition is well-formed if and
only if for each trace $T_i$, $T_i(\delta)\rightarrow \gamma$ such
that $\Gamma \sqsubseteq \gamma$, where $\Delta \sqsubseteq \delta$.
\label{def:well-formed}
\end{definition}

\begin{theorem}
Basic Composition $a\sqsubseteq a_1;a_2$ is well-formed if $\Delta
\sqsubseteq \Delta_1$, $\Delta_2\sqsubseteq \Gamma_1$, and $\Gamma
\sqsubseteq \Gamma_2$.\label{theorem:basic}
\end{theorem}
\noindent {\bf Proof.~}\\
\begin{minipage}{6in} 
\begin{center}
\begin{tabular}{llp{7.5cm}p{3.8cm}}
Initial State & Trace &  Proof step & \\
$\Delta \sqsubseteq \delta$ & $T_1 = a_1;a_2$ & $\Delta_1 \sqsubseteq \Delta $ $~\&~$ $ \Delta \sqsubseteq \delta \Rightarrow \Delta_1 \sqsubseteq \delta$ & (by transitivity)  \hfill(\themyCounter)\\
& & $a_1(\delta)\rightarrow \gamma_1 $ $~\&~$ $ \Gamma_1 \sqsubseteq \gamma_1$ & (Definition of $a_1$)\\
& & $\Delta_2 \sqsubseteq \Gamma_1 $ $~\&~$ $ \Gamma_1 \sqsubseteq \gamma_1 \Rightarrow \Delta_2 \sqsubseteq \gamma_1$ & (by transitivity) \hfill(\themyCounter)\\
& & $a_2(\gamma_1)\rightarrow \gamma_2 $ $~\&~$ $ \Gamma_2 \sqsubseteq \gamma_2$ & (Definition of $a_2$)\\
& & $\Gamma \sqsubseteq \Gamma_2 $ $~\&~$ $ \Gamma_2 \sqsubseteq \gamma_2 \Rightarrow \Gamma \sqsubseteq \gamma_2$ & (by transitivity) \hfill(\themyCounter)\\
\multicolumn{4}{l}{From (1), $a_1$ can be performed in the initial state.}\\
\multicolumn{4}{l}{From (2), $a_2$ can be performed after $a_1$.}\\
\multicolumn{4}{l}{From (1), (2) and (3), $T_1$ is a valid action trace.}\\
\multicolumn{4}{l}{$T_1$ is a valid trace; therefore, $a\sqsubseteq a_1;a_2$ is well-formed.}\hfill$\Box$\\
\end{tabular}
\end{center}
\end{minipage}\\

\begin{theorem}
Basic and Strict Composition $a\sqsubseteq a_1 \vee_s a_2$ is
well-formed if $\Delta_1 \sqcap \Delta_2 \sqsubseteq \Delta$,
$\Gamma \sqsubseteq \Gamma_1$, and $\Gamma \sqsubseteq \Gamma_2$.
\label{theorem:basicstrictchoice}
\end{theorem}
\noindent {\bf Proof.~}\\
\begin{minipage}{6in} 
\begin{center}
\begin{tabular}{llp{8cm}p{3.8cm}}
Initial State & Trace &  Proof step & \\
$\Delta \sqsubseteq \delta$ & $T_1 = a_1$ & $\Delta_1 \sqcap \Delta_2 \sqsubseteq \Delta \Rightarrow \Delta_1 \sqsubseteq \Delta$& (by set inclusion)\\
& & $\Delta_1 \sqsubseteq \Delta $ $~\&~$ $ \Delta \sqsubseteq \delta \Rightarrow \Delta_1 \sqsubseteq \delta$ & (by transitivity) \hfill(\themyCounter) \\
& & $a_1(\delta)\rightarrow \gamma_1 $ $~\&~$ $ \Gamma_1 \sqsubseteq \gamma_1$&  (Definition of $a_1$)  \\
& & $\Gamma \sqsubseteq \Gamma_1 $ $~\&~$ $ \Gamma_1 \sqsubseteq \gamma_1 \Rightarrow \Gamma \sqsubseteq \gamma_1$& (by transitivity) \hfill(\themyCounter)\\
& $T_2 = a_2$ & $\Delta_1 \sqcap \Delta_2 \sqsubseteq \Delta \Rightarrow \Delta_2 \sqsubseteq \Delta$& (by set inclusion)\\
& & $\Delta_2 \sqsubseteq \Delta $ $~\&~$ $ \Delta \sqsubseteq \delta \Rightarrow \Delta_2 \sqsubseteq \delta$ & (by transitivity) \hfill(\themyCounter) \\
& & $a_2(\delta)\rightarrow \gamma_2 $ $~\&~$ $ \Gamma_2 \sqsubseteq \gamma_2$& (Definition of $a_2$) \\
& & $\Gamma \sqsubseteq \Gamma_2 $ $~\&~$ $ \Gamma_2 \sqsubseteq \gamma_2 \Rightarrow \Gamma \sqsubseteq \gamma_2$& (by transitivity) \hfill(\themyCounter)\\
\end{tabular}
\end{center}
\end{minipage}

\begin{minipage}{6in} 
\begin{center}
\begin{tabular}{llp{7.5cm}p{3.8cm}}
\multicolumn{4}{l}{From (4), $a_1$ can be performed in the initial state.}\\
\multicolumn{4}{l}{From (4) and (5), $T_1$ is a valid action trace.}\\
\multicolumn{4}{l}{From (6), $a_2$ can be performed in the initial state.}\\
\multicolumn{4}{l}{From (6) and (7), $T_2$ is a valid action trace.}\\
\multicolumn{4}{l}{$T_1$ and $T_2$ are valid traces; therefore,
$a\sqsubseteq a_1 \vee_s a_2$ is well-formed.}\\
\end{tabular}
\end{center}
\end{minipage}

For all states in $\Delta$, both $a_1$ and $a_2$ can be performed.
Hence, the constraints show that the composition is
strict.\hfill$\Box$

\begin{theorem}
Basic and Strict Composition $a\sqsubseteq a_1 ~\wedge_s~ a_2$  is
well-formed if $\Delta_1 \sqsubseteq \Delta$, $\Delta_2 \sqsubseteq
\Delta$, $\Gamma  \sqsubseteq a_1(a_2(\delta))$ and $\Gamma
\sqsubseteq a_2(a_1(\delta))$.
\label{theorem:basicstrictconjunction}\end{theorem}
\noindent {\bf Proof.~}\\
\begin{minipage}{6in} 
\begin{center}
\begin{tabular}{llp{7.5cm}p{3.8cm}}
Initial State & Trace &  Proof steps & \\
$\Delta \sqsubseteq \delta$ & $T_1 = a_1;a_2$ & $\Delta_1 \sqsubseteq \Delta $ $~\&~$ $ \Delta \sqsubseteq \delta \Rightarrow \Delta_1 \sqsubseteq \delta$& (by transitivity) \hfill(\themyCounter)\\
& & $\Gamma \sqsubseteq a_2(a_1(\delta))$& (Hypothesis) \hfill(\themyCounter)\\ 
 & $T_1 = a_2;a_1$ & $\Delta_2 \sqsubseteq \Delta $ $~\&~$ $ \Delta \sqsubseteq \delta \Rightarrow \Delta_2 \sqsubseteq \delta$& (by transitivity) \hfill(\themyCounter)\\
& & $\Gamma \sqsubseteq a_1(a_2(\delta))$& (Hypothesis) \hfill(\themyCounter)\\ 
\multicolumn{4}{l}{From (8), $a_1$ can be performed in the initial state.}\\
\multicolumn{4}{l}{From (9), $a_2$ can be performed after $a_1$.}\\
\multicolumn{4}{l}{From (8) and (9), $T_1$ is a valid action trace.}\\
\multicolumn{4}{l}{From (10), $a_2$ can be performed in the initial state.}\\
\multicolumn{4}{l}{From (11), $a_1$ can be performed after $a_2$.} \\
\multicolumn{4}{l}{From (10) and (11), $T_2$ is a valid action trace.}\\
\multicolumn{4}{l}{$T_1$ and $T_2$ are valid traces; therefore, $a\sqsubseteq a_1 ~\wedge_s~ a_2$ is a well-formed composition.}\hfill$\Box$\\
\end{tabular}
\end{center}
\end{minipage}\\

\begin{theorem}
Basic and Flexible Composition $a\sqsubseteq a_1 \vee a_2$ is
well-formed if $ \Delta_1 \sqcup \Delta_2 \sqsubseteq \Delta$,
$\Delta_1 \sqcap \Delta \neq \{ \} $, $\Delta_2 \sqcap \Delta \neq
\{ \}$,  $ \Gamma \sqsubseteq \Gamma_1 $, and $ \Gamma \sqsubseteq
\Gamma_2$. \label{theorem:basicflexiblechoice}\end{theorem}
\noindent {\bf Proof.~}\\
\begin{minipage}{6in} 
\begin{center}
\begin{tabular}{llp{8cm}p{3.8cm}}
Initial State & Trace &  Proof steps & \\
$\Delta \sqsubseteq \delta$ &  & $\Delta_1 \sqcup \Delta_2 \sqsubseteq \Delta $ $~\&~$ $ \Delta \sqsubseteq \delta \Rightarrow \Delta_1 \sqcup \Delta_2 \sqsubseteq \delta$& (by transitivity) \hfill(\themyCounter)\\
 & &$\Delta_1 \sqcap \Delta \neq \{ \} $ $~\&~$ $ \Delta_2 \sqcap \Delta \neq \{ \} $ & (Hypothesis) \hfill(\themyCounter)\\
$\Delta_1 \sqsubseteq \delta$ & $T_1 = a_1$ & $a_1(\delta)\rightarrow \gamma_1 $ $~\&~$ $ \Gamma_1 \sqsubseteq \gamma_1$ & (Definition of $a_1$) \hfill(\themyCounter) \\
$\Delta_2 \sqsubseteq \delta$ & $T_2 = a_2$ & $a_2(\gamma_1)\rightarrow \gamma_2 $ $~\&~$ $ \Gamma_2 \sqsubseteq \gamma_2$ & (Definition of $a_2$) \hfill(\themyCounter) \\
\multicolumn{4}{l}{From (12), Initial state $\delta$ is in either $\Delta_1$, $\Delta_2$ or both. Therefore, at least one of $a_1$ and $a_2$ can}\\
\multicolumn{4}{l}{be performed.} \\
\multicolumn{4}{l}{From (13), There are states in $\Delta$ which provide a choice between $a_1$ and $a_2$.}\\
\multicolumn{4}{l}{From (12) and (14), $T_1$ is a valid trace.} \\
\multicolumn{4}{l}{From (12) and (15), $T_2$ is a valid trace.}\\
\multicolumn{4}{l}{$T_1$ and $T_2$ are valid traces; therefore,
$a\sqsubseteq a_1 \vee a_2$ is well-formed.} \hfill$\Box$
\end{tabular}
\end{center}
\end{minipage}\\

\begin{theorem}
Basic and Flexible Composition $a\sqsubseteq a_1 \wedge a_2$ is
well-formed if $ \Delta_1 \sqcup \Delta_2 \sqsubseteq \Delta$, and
for all $\delta \in \Delta$ if $\Delta_1 \sqsubseteq \delta$ then $
\Delta_2 \sqsubseteq a_1(\delta)$ and $\Gamma \sqsubseteq
a_2(a_1(\delta))$, and if $ \Delta_2 \sqsubseteq \delta$ then
$\Delta_1 \sqsubseteq a_2(\delta)$ and $\Gamma \sqsubseteq
a_1(a_2(\delta))$.\label{theorem:basicflexibleconjunction}
\end{theorem}
\noindent {\bf Proof.~}\\
\begin{minipage}{6in} 
\begin{center}
\begin{tabular}{llp{7.5cm}p{3.8cm}}
Initial State & Trace &  Proof steps & \\
$\Delta \sqsubseteq \delta$ &  & $\Delta_1 \sqcup \Delta_2 \sqsubseteq \Delta ~~\&~~ \Delta \sqsubseteq \delta \Rightarrow \Delta_1 \sqcup \Delta_2 \sqsubseteq \delta$& (by transitivity) \hfill(\themyCounter) \\
$\Delta_1 \sqsubseteq \delta$ & $T_1 = a_1;a_2$ & $\Delta_2 \sqsubseteq a_1(\delta) $& (Hypothesis) \hfill(\themyCounter)\\
 & & $\Gamma \sqsubseteq a_2(a_1(\delta))$ & (Hypothesis) \hfill(\themyCounter)\\
$\Delta_2 \sqsubseteq \delta$ & $T_2 = a_2;a_1$ & $\Delta_1 \sqsubseteq a_2(\delta) $& (Hypothesis) \hfill(\themyCounter)\\
 & & $\Gamma \sqsubseteq a_1(a_2(\delta))$ & (Hypothesis) \hfill(\themyCounter)\\
\multicolumn{4}{l}{From (16), Initial state $\delta$ is in either
$\Delta_1$, $\Delta_2$ or both. Therefore, at least one of $a_1$ or}\\
\multicolumn{4}{l}{$a_2$ can be performed. This constraint preserves the semantics of flexible composition.} \\
\multicolumn{4}{l}{From (17), $a_2$ can be performed after $a_1$.} \\
\multicolumn{4}{l}{From (17) and (18), $T_1$ is a valid action trace.} \\
\multicolumn{4}{l}{From (19), $a_1$ can be performed after $a_2$.} \\
\multicolumn{4}{l}{From (19) and (20), $T_2$ is a valid action trace.} \\
\multicolumn{4}{l}{$T_1$ and $T_2$ are valid traces; therefore,
$a\sqsubseteq a_1 \wedge a_2$ is a well-formed composition.}
\hfill$\Box$
\end{tabular}
\end{center}
\end{minipage}\\

\begin{theorem}
Advanced Composition $a\sqsubseteq a_1;[\Delta']a_2$ is well-formed
under following conditions: 1) $\Delta_1 \sqsubseteq \Delta$, 2) if
$\Delta' \not \sqsubseteq a_1(\delta)$ then $\Gamma$ must be refined
by $a_1(\delta)$, i.e., $\Delta' \not \sqsubseteq a_1(\delta)
\Rightarrow \Gamma \sqsubseteq a_1(\delta)$, else performing action
$a_2$ after $a_1$ must lead to a state in $\Gamma$, i.e., $\Delta'
\sqsubseteq a_(\delta) \Rightarrow \Gamma \sqsubseteq
a_2(a_1(\delta))$. \label{theorem:advancedseq} \end{theorem}
\noindent {\bf Proof.~}\\
\begin{minipage}{6in} 
\begin{center}
\begin{tabular}{llp{7.5cm}p{3.8cm}}
Initial State & Trace &  Proof steps & \\
$\Delta \sqsubseteq \delta$ & $T_1 = a_1$ & $\Delta_1 \sqsubseteq \Delta $ $~\&~$ $ \Delta \sqsubseteq \delta \Rightarrow \Delta_1 \sqsubseteq \delta $& (by transitivity) \hfill(\themyCounter) \\
& & $\Delta' \not \sqsubseteq a_1(\delta)$ $~\&~$ $\Gamma \sqsubseteq a_1(\delta)$ & (Hypothesis) \hfill(\themyCounter)\\
& $T_2 = a_1;a_2$ &  $\Delta_1 \sqsubseteq \Delta $ $~\&~$ $ \Delta \sqsubseteq \delta \Rightarrow \Delta_1 \sqsubseteq \delta $& (by transitivity) \hfill(\themyCounter)\\
& & $\Delta' \sqsubseteq a_1(\delta)$ $~\&~$ $\Gamma \sqsubseteq a_2(a_1(\delta))$& (Hypothesis) \hfill(\themyCounter)\\
\multicolumn{4}{l}{From (21), $a_1$ can be performed in initial state.}\\
\multicolumn{4}{l}{From (21) and (22), $T_1$ is a valid trace, when $a_2$ can be ignored.}\\
\multicolumn{4}{l}{From (23), $a_1$ can be performed in initial state.}\\
\multicolumn{4}{l}{From (23) and (24), $T_2$ is a valid trace, when $a_2$ can be performed after $a_1$. }\\
\multicolumn{4}{l}{$T_1$ and $T_2$ are valid traces; therefore,
$a\sqsubseteq a_1;[\Delta']a_2$ is a well-formed composition.}
\hfill$\Box$
\end{tabular}
\end{center}
\end{minipage}\\

\begin{theorem}
Advanced and Strict Composition $a\sqsubseteq
a_1\wedge_s[\Delta']a_2$ is well-formed if the following constraints
are satisfied: 1) if $\Delta_1 \sqsubseteq \Delta$ and $\Delta
\sqcap \Delta' \not \sqsubseteq \delta$ then
 $\Gamma \sqsubseteq a_1(\delta)$,
2) if $\Delta \sqcap \Delta' \sqsubseteq \delta$ then
  $\Delta_1 \sqsubseteq \delta$ and $\Delta' \sqsubseteq
  a_1(\delta)$ and $\Gamma \sqsubseteq a_2(a_1(\delta))$, and
3) if $\Delta \sqcap \Delta' \sqsubseteq \delta$ then
 $\Delta_1 \sqsubseteq a_2(\delta)$ and $\Gamma \sqsubseteq
 a_1(a_2(\delta))$, and 4) $\Delta \sqcap
\Delta' \neq \{\}$\\
\label{theorem:advancedstrict} \end{theorem}
\noindent {\bf Proof.~}\\
\begin{minipage}{6in} 
\begin{center}
\begin{tabular}{llp{7.5cm}p{3.8cm}}
Initial State & Trace &  Proof steps & \\

$\Delta \sqcap \Delta' \not \sqsubseteq \delta$ & $T_1 = a_1$ &
$\Delta_1 \sqsubseteq \Delta $ $~\&~$ $ \Delta \sqsubseteq \delta
\Rightarrow \Delta_1 \sqsubseteq \delta$ &
(by transitivity) \hfill(\themyCounter)  \\

& & $\Gamma \sqsubseteq a_1(\delta)$ & (Hypothesis) \hfill(\themyCounter)  \\

$\Delta \sqcap \Delta' \sqsubseteq \delta$ & $T_2 = a_1;a_2$ &
$\Delta_1 \sqsubseteq \Delta $ $~\&~$ $ \Delta \sqsubseteq \delta
~\Rightarrow~
\Delta_1 \sqsubseteq \delta$ & (by transitivity) \hfill(\themyCounter) \\

& & $\Delta_2 \sqsubseteq a_1(\delta)$ & (Hypothesis)  \hfill(\themyCounter)\\

& & $\Gamma \sqsubseteq a_2(a_1(\delta))$ & (Hypothesis)  \hfill(\themyCounter)\\

& $T_3 = a_2;a_1$ & $\Delta \sqcap \Delta' \sqsubseteq \delta
\Rightarrow \Delta' \sqsubseteq \delta$ &  (by set inclusion)  \hfill(\themyCounter)\\

& & $ \Delta_2 \sqsubseteq \Delta' $ $~\&~$ $ \Delta' \sqsubseteq
\delta ~\Rightarrow~ \Delta_2 \sqsubseteq \delta$ &   (by transitivity) \hfill(\themyCounter)\\

& & $\Delta_1 \sqsubseteq a_2(\delta)$& (Hypothesis)  \hfill(\themyCounter)  \\

& & $\Gamma \sqsubseteq a_1(a_2(\delta))$ & (Hypothesis)  \hfill(\themyCounter) \\

\multicolumn{4}{l}{From (25), $a_1$ can be performed in initial state.}\\
\multicolumn{4}{l}{From (25) and (26), $T_1$ is a valid trace.}\\
\multicolumn{4}{l}{From (27), $a_1$ can be performed in initial state}\\
\multicolumn{4}{l}{From (28), $a_2$ can be performed after $a_1$. }\\
\multicolumn{4}{l}{From (27), (28), and (29), $T_2$ is a valid trace.}\\
\multicolumn{4}{l}{From (31), $a_2$ can be performed in initial state.}\\
\multicolumn{4}{l}{From (32), $a_1$ can be performed after $a_2$}\\
\multicolumn{4}{l}{From (31), (32), and (33), $T_3$ is a valid trace.}\\
\multicolumn{4}{l}{The constraint $\Delta \sqcap \Delta' \neq \{\}$ ensures that the trace $a_2;a_1$ exists. This preserves the semantics of a}\\
\multicolumn{4}{l}{strict conjunction.}\\
\multicolumn{4}{l}{$T_1$, $T_2$, and $T_3$ are valid traces;
therefore, $a\sqsubseteq a_1\wedge_s[\Delta']a_2$ is a well-formed
composition.}\hfill$\Box$
\end{tabular}
\end{center}
\end{minipage}\\

\begin{theorem}
Advanced and Flexible Composition $a\sqsubseteq
a_1\wedge[\Delta']a_2$ is well-formed under following conditions: 1)
$\Delta_1 \sqcup \Delta' \sqsubseteq  \Delta$, 2) if $\Delta_1
\sqsubseteq \delta$ and $\Delta' \sqsubseteq a_1(\delta)$ then
$\Gamma \sqsubseteq a_2(a_1(\delta))$, 3) if $\Delta_1 \sqsubseteq
\delta$ and $\Delta' \not \sqsubseteq \delta$ and $\Delta' \not
\sqsubseteq a_1(\delta)$ then $\Gamma \sqsubseteq a_1(\delta)$, 4)
if $\Delta' \sqsubseteq \delta$ and $\Delta_1 \sqsubseteq
a_2(\delta)$ then  $\Gamma \sqsubseteq a_1(a_2(\delta))$.
\label{theorem:advancedflexible} \end{theorem}
\noindent {\bf Proof.~}\\
\noindent
\begin{minipage}{6in} 
\begin{center}
\begin{tabular}{llp{6.3cm}p{3.8cm}}
Initial State & Trace &  Proof steps & \\
$\Delta \sqsubseteq \delta$ & & $\Delta_1 \sqcup \Delta' \sqsubseteq \Delta $ $~\&~$ $  \Delta \sqsubseteq \delta \Rightarrow \Delta_1\sqcup\Delta' \sqsubseteq \delta$ & (by transitivity) \hfill(\themyCounter)\\
$ \Delta_1 \sqsubseteq \delta$ & $T_1 = a_1;a_2$ & If $\Delta' \sqsubseteq a_1(\delta)$, & (Case)\\
& & $\Delta_2\sqsubseteq\Delta'~\&~\Delta'\sqsubseteq a_1(\delta) \Rightarrow \Delta_2\sqsubseteq a_1(\delta)$ & (by transitivity) \hfill(\themyCounter) \\
& & $\Gamma \sqsubseteq a_2(a_1(\delta))$ & (Hypothesis) \hfill(\themyCounter) \\

& $T_2 = a_1$ & If $\Delta' \not \sqsubseteq a_1(\delta)$ & (Case) \\ 
& & $\Gamma \sqsubseteq a_1(\delta)$ & (Hypothesis) \hfill(\themyCounter)\\

$\Delta' \sqsubseteq \delta$& $T_3 = a_2;a_1$ & $\Delta_2 \sqsubseteq \Delta' ~\&~ \Delta' \sqsubseteq \delta \Rightarrow \Delta_2 \sqsubseteq \delta$ & (by transitivity) \hfill(\themyCounter) \\
& & $\Delta_1 \sqsubseteq a_2(\delta)$ & (Hypothesis)\hfill(\themyCounter)\\
& & $\Gamma \sqsubseteq a_1(a_2(\delta))$ & (Hypothesis)\hfill(\themyCounter)\\
\multicolumn{4}{l}{From (34), Initial state $\delta$ is in either $\Delta_1$, $\Delta'$ or both. Therefore, at least one of $a_1$ or $a_2$ can}\\
\multicolumn{4}{l}{be performed.}\\
\multicolumn{4}{l}{From (35), $a_2$ can be performed after $a_1$.}\\
\multicolumn{4}{l}{From (35) and (36), $T_1$ is a valid trace.}\\
\multicolumn{4}{l}{From (37), $T_2$ is a valid trace.}\\
\multicolumn{4}{l}{From (38), $a_2$ can be performed in the initial state.}\\
\multicolumn{4}{l}{From (39), $a_1$ can be performed after $a_2$.}\\
\multicolumn{4}{l}{From (38), (39) and (40), $T_3$ is a valid trace.}\\
\multicolumn{4}{l}{$T_1$, $T_2$, and $T_3$ are valid traces;
therefore, $a\sqsubseteq a_1\wedge[\Delta']a_2$ is a well-formed
composition.}\hfill$\Box$
\end{tabular}
\end{center}
\end{minipage}\\

\begin{theorem}(Well-formed Complex Composition)\\
Let $a_1:\Delta_1\rightarrow\Gamma_1$ and
$a:\Delta\rightarrow\Gamma$ be composite actions $a_1 \sqsubseteq
a_3 \oplus a_4$ and $a \sqsubseteq a_1 \oplus a_2$ respectively. An
action composition $a \sqsubseteq (a_3 \oplus a_4) \oplus a_2$ is a
well-formed composition if the action compositions $a \sqsubseteq
a_1 \oplus a_2$ and $a_1 \sqsubseteq a_3 \oplus a_4$ are
well-formed.
 \label{thm:complexcomposition}
\end{theorem}
{\noindent {\bf Proof.~}} If $a_1$ is refined by the composition
$a_3\oplus a_4$, all traces of the $a_3\oplus a_4$ must be valid.
Also, it must be possible to perform $a_3 \oplus a_4$ for all states
in $\Delta_1$. From Def.~\ref{def:action-refinement}, if
$a_1\sqsubseteq a_3\oplus a_4$, then for all $\delta \in \Delta_1$,
$a_1(\delta)\rightarrow \gamma_1$, such that, \\
\indent $(a_3\oplus a_4)(\delta)\rightarrow \gamma_{34}$ and $\gamma_1 \sqsubseteq\gamma_{34}$\hfill({\themyCounter)}\\

\noindent Similarly, for all $\delta \in \Delta$,
$a(\delta)\rightarrow\gamma$, such that \\
$(a_1\oplus a_2)(\delta)\rightarrow \gamma_{12}$ and $\gamma
\sqsubseteq \gamma_{12}$\hfill{(\themyCounter)}

\noindent Let $a_{34}=a_3\oplus a_4$ be a state transforming
function $a_{34}:\Delta_{34}\rightarrow\Gamma_{34}$. The action
composition $a\sqsubseteq (a_1\oplus a_2)$ is well-formed if all
traces of $a_1\oplus a_2$ are valid even after substitution of $a_1$
with
$a_{34}$. We now prove validity of each possible trace. \\

\noindent
Case 1:{\em If $a_1;a_2$ is a valid trace then $a_{34};a_2$ is a valid trace.}\\
If $a_1;a_2$ is a valid trace then $\exists~ \delta \in \Delta$,
such that $\Delta_1 \sqsubseteq \delta$ \\
and $a_1(\delta)\rightarrow\gamma_1$, such that $\Delta_2\sqsubseteq
\gamma_1$\hfill{(\themyCounter)}\\
and $a_2(\gamma_1)\rightarrow\gamma_2$, such that $\Gamma\sqsubseteq
\gamma_2$\hfill{(\themyCounter)}

\noindent
From~(41) and (43), we get, \\
\indent $\Delta_2 \sqsubseteq \gamma_1 \sqsubseteq \gamma_{34}$ or
$\Delta_2\sqsubseteq \gamma_{34}$ (by transitivity)
\hfill{(\themyCounter)}\\

\noindent Let $a_2(\gamma_{34})\rightarrow\gamma_{234}$, then from
(44) and (45), we get,\\
\indent  $\gamma_2 \sqsubseteq \gamma_{234}$ (by
monotonicity)\hfill{(\themyCounter)} \\

\noindent From (44) and (46), $\Gamma \sqsubseteq \gamma_{234}$ (by
transitivity). Hence, $a_{34};a_2$ is a valid trace. \\

\noindent
Case 2: {\em If $a_2;a_1$ is a valid trace then $a_2;a_{34}$ is a valid trace.}\\
Reasoning is similar to Case 1. \\

\noindent Case 3: {\em If $a_1$ is a valid trace then $a_{34}$ is a valid trace.}\\
If $a_1$ is a valid trace, $\exists ~\delta \in \Delta$ such that,
$a_1(\delta)\rightarrow \gamma_1$ and $\Gamma \sqsubseteq
\gamma_1$\hfill{(\themyCounter)}\\
\noindent From (41) and (47), $\Gamma \sqsubseteq \gamma_{34}$ (by
transitivity). Hence $a_{34}$ is a valid trace. \\

\noindent Case 4: Refinement of $a_1$ does not effect the trace
$a_2$. \hfill$\Box$

Now we give an example of an action composition.
\begin{example}  Let  {\ttfamily InstallFirewall},   {\ttfamily InstallAntiVirus}, and
 {\ttfamily Protect} be types of actions. Let  {\ttfamily a = Protect((target,\$x))} be a
restricted subclass of class  {\ttfamily Protect}, where {\ttfamily
\$x} is an object variable representing objects that satisfy the
predicates  {\ttfamily type(\$x,Computer)}, and {\ttfamily
owner(\$x,Alice)}. Composition of  {\ttfamily a}
may be described as follows: \\
 {\ttfamily Protect((target,\$x))} $\sqsubseteq$
 {\ttfamily InstallFirewall((target,\$x))} $~ \wedge ~$ \\
\indent \indent \indent \indent \indent \indent \indent
 {\ttfamily [\$x((os,Windows))]InstallAntiVirus((target,\$x))}\\
This composition is an advanced composition using the conjunction
operator. The sub-action  {\ttfamily Install}-{\ttfamily AntiVirus} must be
performed if the operating system is Windows. Otherwise the user may
choose not to perform this action. \label{eg:actionrefinement}
\end{example}

\section{Policy Specification Language}
\label{sec:language} In this section we briefly describe our
approach to incorporate action refinement in authorization policies.
For this we extend the Flexible Authorization Framework
(FAF)~\cite{Jajodia01} to express obligations,
dispensations, and refinement.
FAF is a logic-based framework to express authorization requirements.
Access control permissions or denials are derived by a sequence of applications
of the authorization rules.  These sequence include the propagation, the conflict
resolution, the decision, and the integrity modules.  In addition, it is ensured
that every access request is either granted or denied, therefore ensuring completeness
of the authorization policy.

In our work, we provide extension of FAF, while preserving its 
properties with respect to completeness and decidability. 
 Our extensions, that include predicates to express obligations,
dispensations, and refinements in FAF will preserve the 
properties of locally stratified logic program.
First, we give a brief overview of FAF.  The FAF syntax is built from constants, variables, and
predefined predicates.  The constants and variables range over authorization objects, subjects,
actions, and roles.  FAF includes the following predicates:
\begin{itemize}
\item $cando(X_s,X_o,X_a)$
\item $dercando(X_s,X_o,X_a)$
\item $do(X_s,X_o,X_a)$
\item $done(X_s,X_o,X_a,X_t)$
\item $over_{AO}$ and $over_{AS}$ for overriding predicates
\item $error$ for integrity viiolations
\item $AOH$ and $ASH$ for object and subject hierarhies
\end{itemize}
For detailed explanation of these predicates, look at reference~\cite{Jajodia01}.  FAF rules
are stratified by assigning levels to the predicates and requiring that the head predicate's
level is equal or higher than the levels of the predicates in the rule body.
Formal properties of FAF, such as unique stable model and well-founded model, as well as complexity
analysis, are presented in~\cite{Jajodia01}.

In this work, we propose new predicates to express obligation and dispensation requirements. Table~\ref{table:strata} shows the levels of these predicates along with the original FAF
predicates.  First, we start with the formal description of
these concepts.

Regulations often specify obligations as one of their requirements.
In general, we interpret obligations as actions that users are
required to perform to achieve specific goals. 
\begin{definition}(Obligation) \\
Let $A: \Delta \rightarrow \Gamma$ be an action type. An obligation
$o = oblig(s, A, q)$ is defined as a command to subject $s$ to
perform an action of type $A$, such that the condition $q$ is
satisfied. Definition of an obligation is said to be correct if
$\Gamma \sqcap \Gamma_q \neq \{\}$, where $\Gamma_q$ is state space
representing all states in which $q$ is true. Let $\delta$ be the
state of a given system. We say that subject $s$ has satisfied
obligation $O$ if $\Gamma \sqcap \Gamma_q \sqsubseteq \gamma$. If
$\Delta \not \sqsubseteq \delta$, the assumptions made to perform
the action of type $A$ are violated. Violating the assumptions
releases the subject from the obligation. As this is not fault of
the subject, it is considered to have satisfied the obligation.
 \label{def:obligation}
\end{definition}
\begin{definition}(Dispensation)\\
Let $A: \Delta \rightarrow \Gamma$ be an action type. A dispensation
$d = disp(s, A)$ is defined as an exemption given to subject $s$
from performing an action of type $A$.
 \label{def:dispensation}
\end{definition}

Rules in our policy language consists of constants, variables, and
predicates. They are defined as follows:

\begin{enumerate}
\item {\em Constant Symbols:} Every member of
$Obj \cup T \cup U \cup G \cup R \cup A$, where Obj is the set of
objects, T the set of types, U the set of users, G the set of
groups, R the set of roles, A the set of action types.
\item {\em Variable Symbols:} There are seven sets $V_o$, $V_t$ ,
$V_u$, $V_g$ , $V_r$ , $V_{a}$ of variable symbols ranging over the
sets $Obj$, $T$, $U$, $G$, $R$, $A$, respectively. \item {\em
Predicate Symbols:}
    \begin{enumerate}
    \item A 3-ary predicate symbol, {\ttfamily hasObligation}. The first argument is
    a subject term, the second argument is an action term,
    and the third argument is a boolean formula called post-condition.
    \item A 2-ary predicate symbol, {\ttfamily hasDispensation}. The first argument
    is a subject term, and the second argument is an action term.
    \item A 3-ary predicate symbol, {\ttfamily derhasObligation}, with the same arguments
    as {\em hasObligation}. The predicate {\ttfamily derhasObligation} represents obligations
    derived by using logical rules of inference
    (modus ponens plus rules for stratified negation~\cite{Apt88}).
    \item A 2-ary predicate symbol, {\ttfamily derhasDispensation}, with the same arguments
    as {\em hasDispensation}. The predicate {\ttfamily derhasDispensation} represents
    dispensations derived by using logical rules of inference
    (modus ponens plus rules for stratified negation).
    \item A 3-ary predicate symbol, {\ttfamily mustdo}, with the same arguments
    as {\ttfamily hasObligation} and {\ttfamily derhasObligation}. It definitely represents
    the actions that must be performed. Intuitively, {\ttfamily mustdo} enforces the conflict
    resolution and obligation policy.
    \end{enumerate}
\end{enumerate}

In addition, we allow use of {\ttfamily cando, dercando, do, done,
$over_{AS}$, $over_{AO}$, error}, $hie-$, and $rel$ predicates as
defined in FAF. Table~\ref{table:strata} shows the strata of rules
allowed in our policy to represent obligations, dispensations and
their refinement.

\begin{table}[h]
\begin{center}
\begin{minipage}[t]{6in}
\footnotesize
\begin{tabular}{|l|l|l|p{3.2in}|}
\hline

Level & Stratum & Predicate & Rules defining predicate \\

\hline

0 & $S_0$ & hie-predicates & base relations \\
  &     & rel-predicates & base relations \\
  &     & done & base relation \\

\hline

1 & $S_1$ & hasObligation & body may contain done, hie- and rel- literals. \\
  &        & hasDispensation & body may contain done, hie- and rel- literals. \\



%

\hline

2 & $S_2$ & derhasDispensation & body may contain hasObligation,
hasDispensation, derhasDispensation, over, done,
hie- and rel- literals.\\
\hline

3 & $S_3$ & derhasObligation & body may contain hasObligation,
hasDispensation,
derhasObligation, derhasDispensation, over, done, hie- and rel- literals.\\

\hline

4 & $S_4$ & mustdo & body may contain hasObligation,
derhasObligation, hasDispensation, derhasDispensation, done,
hie- and rel- literals.\\

\hline

5 & $S_5$ & cando & body may contain mustdo, done,
hie- and rel- literals.\\

\hline

6 & $S_6$ & dercando & body may contain mustdo, cando, dercando,
 done, hie- and rel- literals.\\

\hline

7 & $S_7$ & do & in the case when head is of the form do(o,s,+a)body
may contain cando, dercando, done,
hie- and rel- literals.\\
\hline

 8 & $S_8 $& do & in the case when head is of the form
do(o,s,-a) body
contains just one literal $\neg$do(o,s,+a).\\

\hline

9 & $S_9$ & error & body may contain mustdo, hasObligation, derhas-
Obligation, hasDispensation, derhasDispensation, do, cando,
dercando, done,
hie- and rel- literals.\\

\hline
\end{tabular}
\normalsize
\end{minipage}
\end{center}
\caption{Obligation and Authorization Specification Strata}
\label{table:strata}
\end{table}

\begin{definition}(Obligation Rule)\\
An {\em obligation rule} is a rule of the form:\\
{\ttfamily hasObligation(s,a,q)} $\leftarrow L_1 \& \ldots \& L_n$\\
 where $s$
is a subject term, $a$ is an obligation action type, $q$ is a
boolean formula composed with $rel$- predicates and $done$ literals,
and $L_1 \& \ldots \& L_n$ are $done$, $hie$- or $rel$- literals.
\label{def:obligationrule}
\end{definition}

\begin{example}  Let us assume that an organization requires
computers to have firewall software installed to be considered safe.
The obligation "Employees must protect computers they own from
unauthorized access" is then modelled by following obligation rule:

{\ttfamily hasObligation(\$s, Protect((target,\$x))},
{\ttfamily hasInstalled(\$x, \$y)} \\
{\ttfamily\&} {\ttfamily type(\$y,Firewall))}
 $\leftarrow$ {\ttfamily type(\$x,Computer)} {\ttfamily \&}
{\ttfamily type(\$s,Employee) \\ {\ttfamily \&} owner(\$x,\$s)}

where $\$x, \$y$, and $\$s$ are variables, {\ttfamily Protect} is a
sub-class of {\ttfamily Action}, {\ttfamily Computer} and {\ttfamily
Employee} are sub-classes of {\ttfamily Object}, {\ttfamily type} is
a $hie$ predicate, and {\ttfamily target}, {\ttfamily hasInstalled},
and {\ttfamily owner} are $rel$ predicates.

Let us assume that the data system contains two Employee objects and
three Computer objects such that the following predicates hold in
the system state:

{\ttfamily type(pc1, Computer)}, {\ttfamily type(emp1, Employee)}\\
\indent {\ttfamily type(pc2, Computer)}, {\ttfamily type(emp2, Employee)}\\
\indent {\ttfamily type(pc3, Computer)}, {\ttfamily owner(emp1, pc1)}\\
\indent {\ttfamily owner(emp2, pc2)}, {\ttfamily owner(emp1, pc3)}

When above obligation rule is evaluated in the data system presented
above, the results of evaluations are {\ttfamily
(\$x=pc1,\$s=emp1)}, {\ttfamily (\$x=pc2,\$s=emp2)}, and {\ttfamily
(\$x=pc3, \$s=emp1)}. Applying the evaluation results to the
obligation rule creates following three obligations:

\noindent
{\ttfamily hasObligation(emp1, Protect((target,pc1)), hasInstalled(pc1, \$y) \\ \indent ~\&~type(\$y,Firewall))} \\
{\ttfamily hasObligation(emp2, Protect((target,pc2)), hasInstalled(pc2, \$y) \\ \indent ~\&~type(\$y,Firewall))} \\
{\ttfamily hasObligation(emp1, Protect((target,pc3)), hasInstalled(pc3, \$y) \\ \indent  ~\&~type(\$y,Firewall))} \\
\end{example}

\begin{definition}(Dispensation Rule)\\
A {\em dispensation rule} is a rule of the form: \\
{\ttfamily hasDispensation(s,a)} $\leftarrow L_1 \& \ldots \& L_n$ \\
where $s$ is a subject term, $a$ is an obligation action type, and
$L_1 \& \ldots \& L_n$ are $done$, $hie$- or $rel$- literals.
\label{def:dispensationrule}
\end{definition}

New obligations and dispensation may be derived from existing
obligations, dispensations, hie- and rel- predicates using inference
rules called derivation rules. For example, a derivation rule can
specify propagation of obligation via subject hierarchy, and
delegation of duties. Definition of dispensation and obligation
derivation rules follow.

\begin{definition}(Dispensation Derivation Rule)\\
A {\em dispensation derivation rule} is a rule of the form: \\
{\ttfamily derhasDispensation(s,a)} $\leftarrow L_1 \& \ldots \& L_n$ \\
where $s$ is a subject term, $a$ is an obligation action type, and
$L_1 \& \ldots \& L_n$ are {\ttfamily hasDispensation,
derhasDispensation, done}, $hie$- or $rel$- literals. All {\ttfamily
derhasDispensation} literals appearing in the body must be positive.
\label{def:derdispensationrule}
\end{definition}

\begin{definition}(Obligation Derivation Rule)\\
An {\em obligation derivation rule} is a rule of the form: \\
\indent {\ttfamily derhasObligation(s,a,q)} $\leftarrow L_1~\&~\ldots~\&~L_n$\\
where $s$ and $a$ are terms of $ST$ and $OA$ respectively, $q$ is a
system state, and $L_1~\&\ldots\&~L_n$ are {\ttfamily hasObligation,
derhasObligation, derhasDispensation, done}, $hie$- or $rel$-
literals. All {\ttfamily derhasObligation} literals appearing in the
body must be positive.\label{def:derivationrule}
\end{definition}

\begin{definition}(Derivation View)\\
A derivation view is a finite set of derivation rules.
\end{definition}

\subsection{Policy Refinement}

We use derivation rules to refine a high-level policy into low-level
policy. Derivation is based on subject hierarchy as in FAF, and
action refinement patterns. A discussion of types of derivation rules
is presented below.

\noindent {\bf A. Derivation via subject-hierarchy}\\
Propagation of obligations and dispensation can be achieved via
subject-hierarchy. Dispensation derivation rules expressing
propagation via subject-hierarchy may have the following form:\\
{\ttfamily derhasObligation(s,a,q) $\leftarrow$
hasObligation(s',a,q) $~\&~$ hie(s,s')} \\
{\ttfamily derhasObligation(s,a,q) $\leftarrow$
derhasObligation(s',a,q) $~\&~$ hie(s,s')} \\
{\ttfamily derhasDispensation(s,a) $\leftarrow$
hasDispensation(s',a) $~\&~$ hie(s,s')  $\&$ \\
\indent\indent\indent\indent    $L_1~ \&\ldots\&~ L_n$} \\
{\ttfamily derhasDispensation(s,a) $\leftarrow$
derhasDispensation(s',a) $~\&~$ hie(s,s')}

where $L_1\&\ldots\&L_n$ are {\ttfamily hasDispensation,
derhasDispensation, done}, $hie$- or $rel$- literals. All {\ttfamily
derhasDispensation} literals appearing in the body must be
positive.

Obligation derivation rules expressing
propagation via subject-hierarchy may have the  following form:\\
{\ttfamily derhasObligation(s,a,q) $\leftarrow$
hasObligation(s',a,q) $\&$ \\
\indent \indent \indent \indent \indent \indent \indent \indent
\indent\indent hie(s,s') $\&$ $L_1~ \&\ldots\&~ L_n$} \\
{\ttfamily derhasObligation(s,a,q) $\leftarrow$
derhasObligation(s',a,q) $\&$ \\
\indent \indent \indent \indent \indent \indent \indent \indent
\indent\indent hie(s,s') $\&$ $L_1~ \&\ldots\&~ L_n$} \\
where $L_1\&\ldots\&L_n$ are {\ttfamily hasObligation,
derhasObligation, derhasDispensation, done}, $hie$- or $rel$-
literals. All {\ttfamily derhasObligation} literals appearing in the
body must be positive.

\begin{example}
Let us assume that a security policy specifies that all employees
have an obligation to protect computers they own. A manager is a
type of an employee. Hence, managers have an obligation to protect
computers they own. This derivation rule is represented as follows: \\

\noindent
{\ttfamily derhasObligation(\$$s_2$, Protect(target \$x),q) $\leftarrow$ \\
\indent hasObligation(\$$s_1$, Protect(target \$x), q) \&
\\ \indent isa(Manager, Employee) \& type(\$$s_1$,Employee) \&
type(\$$s_2$,Manager) \& \\
\indent owns(\$$s_1$,N1) \& type(N1, Computer)}
\end{example}

\noindent
{\bf B. Derivation via action refinement}\\
New obligation rules and dispensation rules may be derived from a
high-level obligation or dispensation rule, by substituting the
action in high-level rule with its sub-actions as specified in
refinement pattern. We now discuss construction of derivation rules
based on basic and strict action composition operators.

Let {\ttfamily hasObligation(s,a,q)}$\leftarrow L_1 ~\&~ \ldots ~\&~
L_n$ be an obligation rule, and let $a_1: \Delta_1 \rightarrow
\Gamma_1$ and $a_2: \Delta_2 \rightarrow \Gamma_2$ be the
sub-actions of $a: \Delta \rightarrow \Gamma$. Then the given
obligation rule can be
refined into obligation rules for sub-actions as described below. \\

\noindent {\bf B.1 Distribution over sequence operator}\\ Let $a
\sqsubseteq a_1 ; a_2$ be the refinement pattern for action of type
$a$. An obligation rule to perform action $a$ can be refined into
two obligations to perform sub-actions $a_1$ and $a_2$ with rules
of following form:\\
\noindent
{\ttfamily derhasObligation(s,$a_1$,$q_1$)} $\leftarrow L_1 ~\&~ \ldots~\&~ L_n$ \\
{\ttfamily derhasObligation(s,$a_2$,$q$)} $\leftarrow$ {\ttfamily done($a_1$)  \& hasObligation(s,a,q)}\\
where $q_1 =  \Gamma_1 \cap \Delta_2$. We constrain the post-condition
of first obligation action $a_1$ to satisfy pre-conditions required
to perform second obligation action $a_2$.

\noindent {\bf B.2 Distribution over choice operator}\\
Let $a \sqsubseteq a_1 \vee a_2$ be the refinement pattern for action
of type $a$. Let $R$ be the rule that derives obligation to perform $a$.
We know that if either $a_1$ or $a_2$ is performed the obligation
is satisfied. Therefore, an obligation rule to perform action $a$
can be refined into either of the following two obligation rules $R_1$
and $R_2$. Application of this refinement pattern to a policy $P$
generates two refined policies $P_1$ and $P_2$. The rule
rule $R$ in $P$ is substituted with $R_1$ and $R_2$ to generate
$P_1$ and $P_2$ respectively.

\noindent
$R_1$: \indent {\ttfamily derhasObligation(s,$a_1$,q) $\leftarrow$ $L_1~\&~\ldots~\&~L_n~\&~\neg$done($a_2$)}  \\
$R_2$: \indent {\ttfamily derhasObligation(s,$a_2$,q) $\leftarrow L_1~\&~\ldots~\&~L_n~\&~\neg$done($a_1$)}\\

\noindent {\bf B.3 Distribution over conjunction operator}\\
When an action $a$ is refined by an action composition of form
$a_1 \wedge a_2$, we refine the policy in two steps. First, we
substitute $a_1 \wedge a_2$ with the composition $a_3 \vee a_4$,
where $a_3 \sqsubseteq a_1;a_2$ and $a_4 \sqsubseteq a_2;a_1$.
This allows us to apply action refinement mechanism for choice operator
as we described above. In second step, we refine actions $a_3$
and $a_4$ in resulting policies using the action refinement
mechanism for sequence operator.


\subsection{Deriving Authorizations}

Security policies may also contain authorization rules in addition
to obligation and dispensation rules. Moreover, the policy
refinement mechanism presented in the previous section can be
extended by adding rules that derive permissions and prohibitions from
predicates defined in obligation specification strata. In this
section, we examine authorization rules that may contain obligations
and dispensations.

From the perspective of refining a security policy, an obligation to
perform an action suggests that the subject must have permission to
perform or execute the obligation action.

\begin{definition}(Authorization Rule)\\
An {\em authorization rule} is a rule of the form: \\
\indent \indent \indent {\ttfamily cando(o,s,<sign>a)} $\leftarrow L_1 \& \ldots \& L_n$ \\
where $s$ is a subject term, $a$ is a signed action type, $sign$ is
either + or -, and $L_1 \& \ldots \& L_n$ are {\ttfamily mustdo,
done}, $hie$- or $rel$- literals. \label{def:authorizationrule}
\end{definition}

\begin{example} Suppose an obligation decision rule is derived
saying that subject $s$ is required to encrypt an object $x$. To be
able to fulfill the obligation $s$ must have permission to execute
$Encrypt$ action or function. \\
{\ttfamily cando(Encrypt((target,x)), s, +execute)} $\leftarrow$\\
\indent \indent \indent{\ttfamily mustdo(s, Encrypt((target,x)), q)}
\label{eg:derivecando}
\end{example}

Obligation to perform an action can also imply prohibition to
perform certain actions. Prohibitions are represented by
authorization rules specifying a - sign for the action.

\begin{example}
Suppose subject $s$ has an obligation to encrypt email messages that
contain confidential messages. To ensure compliance to this policy
rule, the policy-refinement procedure can add a rule disallowing $s$
to send email if its contents are confidential. Such a rule may be
expressed as
follows:\\
{\ttfamily cando(sendEmail((message,x)),s,-execute) $\leftarrow$ \\
\indent\indent  mustdo(s, Encrypt((target,x)),q) \& type(x,
EmailMessage) \& \\
\indent\indent  messagetype(x, PlainText) \&
hasClassification(x,Confidential)} \label{eg:deriveprohibition}
\end{example}

To perform an obligation action, the subject $s$ may need
permissions on objects accessed by the obligation action. Objects
that are accessed but not modified are described by {\ttfamily
instrument} property of the class {\ttfamily Action}. Objects that
are accessed and modified by an action are described by {\ttfamily
resource} property of the class {\ttfamily Action}. An obligation to
perform an action can be refined into authorization rules for
instrument and resource objects as shown below: \\
 {\ttfamily cando(\$r, s, +modify) $\leftarrow$ mustdo(s,a,q) \&
 resource(a,\$r)}\\
 {\ttfamily cando(\$i, s, +read) $\leftarrow$ mustdo(s,a,q) \& instrument(a,\$i)}

Authorization derivation in this framework have definition same as
in FAF~{\cite{Jajodia01}}. It is given below to provide complete
description of this framework.

\begin{definition}(Authorization Derivation Rule)\\
An {\em authorization derivation} rule is of the form:\\
\indent\indent\indent {\ttfamily dercando(o,s,$<$sign$>$a) $ \leftarrow  L_1 \& \ldots \& L_n$}\\
where $o$ is an object term, $s$ is a subject term,  $a$ is an
action term,  sign is either + or -, and $L_1, \ldots , L_n$ are
either cando, over, dercando, done, hie-, or rel literals. All
dercando-literals appearing in the body of a derivation rule must be
positive. \label{def:authorizationderivationrule}
\end{definition}

Definition of authorization decision rules in our policy refinement
framework is different than that in FAF. FAF uses a closed policy
and creates a prohibition for all actions that are not explicitly
permitted. However, in policy refinement we assume that the
refinement of high-level policy may not generate all the positive
authorization rules that may be present in the low-level security
policy. We do require that all negative authorization rules generated
by policy refinement must be present in the low-level security
policy. We assume that the high-level policy does not contain
positive authorization rules, and the low-level policy may not override
positive authorizations derived from the high-level policy.

\begin{definition}(Authorization Decision Rule)\\
An {\em authorization decision} rule is of the form:\\
\indent\indent\indent {\ttfamily do(o,s,$<$sign$>$a) $\leftarrow L_1 \& \ldots \& L_n$}\\
where $o$ is an object term, $s$ is a subject term,  $a$ is an
action term,  sign is either + or -, and $L_1, \ldots , L_n$ are
either cando, dercando, done, hie-, or rel literals.
\label{def:authorizationdecisionrule}
\end{definition}

\subsection{Conflict Resolution}

Policy refinement must lead to decision whether a subject has an
obligation to perform an action or not. However, policies may
generate conflicting rules. For example, a subject may have an
obligation to perform an action $a$ and can also have a dispensation
for action $a$ at the same time. Conflict resolution rules are added
to deal with such situations.

A conflict resolution rule expressing that dispensations take
precedence can be of following form: \\
{\ttfamily derhasDispensation(s,a) $\leftarrow$
  hasDispensation(s,a)$~\&~$ \\
  \indent\indent\indent\indent\indent\indent\indent\indent\indent
  \indent\indent\indent\indent\indent\indent\indent\indent\indent
   hasObligation(s,a,q)  \& $L_1 \& \ldots \& L_n$}

Conflict resolution rules that express obligation takes precedence
are expressed as obligation decision rules
(Def.~\ref{def:obligationdecisionrule})

\begin{definition}(Obligation Decision Rule)\\
An obligation decision rule is a rule of the form \\
\indent {\ttfamily mustdo(s,a,q) $\leftarrow L_1~\&~\ldots~\&~L_n$} \\
where $s$ and $a$ are elements of $S$ and $OA$ respectively, $q$ is
a system state, and $L_1~\&~\ldots~\&~L_n$ are {\ttfamily
hasObligation, derhasObligation, hasDispensation, derhasDispensation,
done}, $hie$- or $rel$- literals and every variable
that appears in any of the $L_i$'s also appears in the head of this
rule. \label{def:obligationdecisionrule}
\end{definition}

Separation of duty requires that for a particular set of actions in
a transaction, no single individual be allowed to execute all
actions within the set. Separation of duty is often enforced with
access control policies.  In the policy refinement model presented
in this work, positive authorizations are derived from obligations.
A user must have permissions to perform actions he is obliged to do
as discussed above. However, the derived permissions must reflect
separation of duty requirements. Hence, the obligation and
dispensation rules must be modeled to handle separation of duties.

For example, if the separation of duties require that actions $a_1$
and $a_2$ must not be performed by the same subject. A subject
obliged to perform $a_2$ must be given dispensation on action $a_1$.
In such cases, an additional obligation derivation rule can be
stated to specify alternate subject that will be required to
perform action $a_1$ and complete the transaction successfully.

\noindent {\ttfamily derhasDispensation(s,$a_1$) $\leftarrow$
derhasObligation(s,$a_1$,$q_1$) $~\&~$} \\
\indent {\ttfamily derhasObligation(s,$a_2$,$q_2$)$~\&~L_1 ~\&~\ldots~\&~ L_n$} \\
\noindent {\ttfamily derhasObligation(s',$a_1$,$q_1$) $\leftarrow$
derhasDispensation(s,$a_1$)$~\&~ L'_1~\&\ldots\&~ L'_n$}\\
where $s'$ is a subject term defined in body of the rule.

In addition, a policy may have modal authorization conflicts, i.e.,
policy refinement may generate both positive and negative
authorizations on same object for a subject. For example, policy
refinement may derive positive authorizations for a subject on
objects required to perform his/her obligations. However, there may
be another rule in the policy prohibiting access to the required
object. In this case, a conflict resolution rule may be defined to
allow the subject to access required objects. In general, conflict
resolution rules for authorizations are modeled as authorization
decision rules (Def.~\ref{def:authorizationdecisionrule}).

\begin{definition}(Decision View)\\
A decision view is a finite set of decision rules.
\end{definition}

\begin{definition}(Integrity Rule)\\
An integrity rule is of the following form:
\begin{center}
{\ttfamily error $\leftarrow$ $L_1\&\ldots\&L_n$}
\end{center}
where $L_1,\ldots,L_n$ are mustdo, hasObligation, derhasObligation,
hasDispensation, derhasDispensation, do, cando, dercando, done,
hie-, and rel- literals.
\end{definition}

\begin{definition}(Policy) \\
\noindent A {\em policy} $P = (R,DS,E)$ is a set of rules $R = H
\cup A \cup M$ characterized by its scope $DS$ and environment $E$,
where $H$ is a set of obligation rules and dispensation rules, $A$
is a set of authorization rules, and $M$ is a set of propagation
rules, conflict resolution rules, and integrity rules. Scope
specifies set of target objects to which the policy is applicable,
and environment specifies compliance verification context
information like date, time, location, subject, etc.
\label{def:policy}
\end{definition}

\begin{example} Let us now consider an example illustrating
application of derivation rules, and decision rules for policy
refinement. Consider the following security policy:

\noindent {\ttfamily hasObligation(\$s, Protect((target, \$x)),
true) $\leftarrow$ \\
\indent\indent\indent\indent type(\$s, Employee) \& owns(\$s,\$x)
                            \& type(\$x, Computer)}

\noindent {\ttfamily hasDispensation(\$s, InstallFirewall((target,
\$x))) $\leftarrow$ \\
\indent\indent\indent\indent type(\$s, Employee) \& owns(\$s,\$x)
                            \& type(\$x,Computer) \& \\
\indent\indent\indent\indent hasRole(\$s, Manager)}

\noindent {\ttfamily mustdo(\$s, \$a,\$q) $\leftarrow$
 derhasObligation(\$s,\$a, \$q) \& \\
\indent\indent\indent\indent $\neg$ derhasDispensation(\$s, \$a) }

\noindent Let the refinement for action {\ttfamily Protect} be
defined by following action composition:

{\ttfamily  Protect((target, \$x)) $\sqsubseteq$
 InstallFirewall((target, \$x)) $\wedge$ \\
 \indent\indent\indent\indent InstallAntiVirus((target, \$x))}

\noindent
Let the following predicates hold in system state:\\
{\ttfamily type(Alice, Employee), hasRole(Alice, Manager), \\
 owns(Alice, NB1), type(NB1, Computer)}

To refine the security policy, we first apply the derivation rules to
derive all predicates in stratum $OS_2$, followed by derivation of
all predicates in stratum $OS_3$, and so on. We first, evaluate the
variables in obligation rules and dispensation rules using the
system state. For above security policy, following rules are
derived after evaluation:

\noindent {\ttfamily hasObligation(Alice, Protect((target, NB1)),
true) $\leftarrow$ \\
\indent\indent\indent type(Alice, Employee)
                     \&  owns(Alice,NB1) \& type(NB1, Computer)}

\noindent {\ttfamily hasDispensation(Alice, InstallFirewall((target,
NB1))) $\leftarrow$ \\
\indent\indent\indent type(Alice, Employee) \& owns(Alice,NB1)
                       \& type(NB1,Computer) \\
\indent\indent\indent  \& hasRole(Alice, Manager)}

We now apply derivation rules, e.g., derivation rules for policy
refinement by action refinement. By refining action protect, we
obtain following rules:

\noindent {\ttfamily derhasObligation(Alice,
InstallFirewall((target, NB1)), true) $\leftarrow$ \\
\indent\indent\indent type(Alice,Employee) \& owns(Alice,NB1) \&
type(NB1, Computer)}

\noindent {\ttfamily derhasObligation(Alice,
InstallAntiVirus((target, NB1)), true) $\leftarrow$ \\
\indent\indent\indent type(Alice, Employee) \& owns(Alice,NB1)
 \& type(NB1, Computer)}

No new predicates can be further derived in this level. Hence, we
now apply the decision rules to obtain predicates in higher stratum.
Since, both predicates
{\ttfamily derhasObligation(Alice,
Install}-{\ttfamily Firewall((target, NB1)), true)} and {\ttfamily
derhasDispensation(Alice, InstallFirewall} {\ttfamily ((target, NB1)))} hold, a
{\ttfamily mustdo} predicate for Alice to perform the action
{\ttfamily InstallFirewall} cannot be derived. However,  a
{\ttfamily mustdo} predicate for {\ttfamily InstallAntiVirus} action
is derived from the following instance of decision rule:

\noindent \ttfamily{mustdo(Alice, InstallAntiVirus((target,
NB1)),true) $\leftarrow$ \\
\indent\indent\indent derhasObligation(Alice,
InstallAntiVirus((target, NB1)),true)  \\
\indent\indent\indent \& $\neg$ derhasDispensation(Alice,
InstallAntiVirus((target, NB1)))}
\end{example}

\section{Compliance}
\label{sec:checking}

To check compliance, we compare a high-level security policy
with a low-level security policy in context of a data system.
The set of {\ttfamily do} and {\ttfamily mustdo} ground predicates
that can be derived from a security policy and a data system is
called ground decision view.

\begin{definition}(Compliance)\\
A low-level policy $P_l$ is {\em compliant} to a higher-level policy
$P_h$ for a given $DS$,  if there exists a $(P_h,DS)_{ref}$, such
that $(P_l,DS) \Rightarrow$ $(P_h,DS)_{ref}$, where $DS$ is the data
system, $(P_l,DS)$ represents the ground decision view of low-level
security policy, and $(P_h,DS)_{ref}$ represents the ground decision
view of refined high-level security policy. We assume that $P_h$ does
not contain any positive authorization rules \label{def:compliance}
\end{definition}

\begin{algorithm}[!hb]
\scriptsize \dontprintsemicolon

\SetKwInOut{Input}{input} \SetKwInOut{Output}{output}

\SetKwData{DataSystem}{DS} \SetKwData{Onotology}{O}
\SetKwData{HLP}{$P_h$} \SetKwData{LLP}{$P_l$}
\SetKwData{Patterns}{RP} \SetKwData{State}{$\sigma$}

\Input{ High-level security policy \HLP, Low-level security policy \LLP, Data System \DataSystem, Refinement Patterns \Patterns, Current State \State}
\Output{true if \LLP and \State are compliant to \HLP, otherwise
false}

\BlankLine

// Generate ground decision view of $P_l$ give a data system $DS$ \;

 Evaluate the variables in \LLP.\;

Instantiate the variables in \LLP to derive ground rules. \;

Apply derivation rules and conflict resolution rules until no new
fact is generated. \;

Apply Integrity rules. If errors are found report that policy \LLP
is inconsistent. \;

\BlankLine
// Generate all ground decision views of $P_h$ given a data system $DS$ \;
// Note that multiple decision views may be derived from \HLP. \;

Evaluate the variables in \HLP.\;

\Repeat{no new fact is generated}{

 Instantiate the variables in \HLP to derive ground rules. \;

Apply derivation rules, and conflict resolution rules until no new
fact is generated. \;

Apply Integrity rules. If errors are found report that policy \HLP
is inconsistent. \; }

 \BlankLine
// Compare ground decision views, which consists of authorization
 obligation decision views. \;

compliant $\leftarrow$ false \;

\ForEach{decision view derived from \HLP}{

    found $\leftarrow$ true \;

    // Compare authorization decision views. \;
    Let $D_h$ be the set of {\ttfamily do} predicates derived from \HLP
    and are applicable in \DataSystem and current state \State. \;

    Let $D_l$ be the set of {\ttfamily do} predicates derived from \LLP
    and are applicable in \DataSystem and current state \State. \;

    \If{$D_h$ $\not \subseteq$ $D_l$}{
        found $\leftarrow$ false \;
    }


    \BlankLine

    // Compare obligation decision views \;

    Let $M_h$ be the set of {\ttfamily mustdo} predicates derived from \HLP
    and are applicable in \DataSystem and current state \State. \;

    Let $M_l$ be the set of {\ttfamily mustdo} predicates derived from \LLP
    and are applicable in \DataSystem and current state \State. \;

    (Note that \LLP may have no means to enforce obligations or \LLP may not
      contain obligations. In such cases, we consider $M_l$ to be empty and check
      for satisfaction of obligations in $M_h$.) \;

      \ForEach{predicate {\ttfamily mustdo(s,a,q)} in $M_h$}{
        Let $e_a$ be the effect of action $a$ asserted by ontology. \;
        Compute $e_a$ by evaluating {\ttfamily effect(a, $e_a$)} given data system \DataSystem \;
        \If{not (({\ttfamily mustdo(s,a,q)} in $M_l$) OR (\State $\Rightarrow q$ and \State $\Rightarrow e_a$))}{
            found $\leftarrow$ false \;
        }
      }

    compliant $\leftarrow$ compliant OR found \;
      \If{compliant}{break\;}

}

\Return compliant \;

\caption{Compliance checking algorithm} 
\label{alg:compliance}
\end{algorithm}

Algorithm~\ref{alg:compliance} describes the steps needed to 
check compliance of a give low-level policy and system state
to a given high-level policy. 
First, the algorithm decision view of low-level security policy. 
The low-level policy is a stratified logic program and can be 
evaluated in polynomial time~\cite{Jajodia01}. We then refine the 
high-level policy. The refinement process can lead to multiple
refinements of high-level policies due to action refinement over
the choice choice operator ($\vee$) and conjunction operator 
($\wedge$). The process of refining low-level policies is analogous 
to a top-down tree traversal, where each internal node of the tree
represents the refinement stage at which an action is refined into a 
composition with choice operator or conjunction operator. The leafs 
of the tree represent derivation of refined policies with atomic actions.
Therefore, the complexity of the policy refinement can be seen
as exponential in terms of height of this evaluation tree, which 
corresponds to number of time action refinement has to be 
applied to reach atomic actions. Finally, the algorithm checks for
compliance by searching for a refinement of high-level policy such 
that all the access control and obligation requirements
specified in the refined policy are satisfied
by the low-level policy or current system state. 

A given refinement of high-level policy and low-level policy can 
also be compared to detect conflicts among them. 
We categorize conflicts between a (high-level) security policy and
system state (low-level policy and object properties) into following
four categories:

\begin{definition}(Modal Authorization Violations)\\
A modal violation occurs when a high level policy has granted
authorization but a low level policy denies authorization.\\
\noindent Let $ R_h$ be a authorization decision rule $do(s,o,-a)
\leftarrow L_1  \& \ldots \& \L_n$ in refinement of high level
policy $P_h$, and $ R_l$ be an authorization decision rule $do(s, o,
+a) \leftarrow L'_1 \& \ldots \& \L'_m$ in low-level policy $P_l$.
$R_h$ and $R_l$ have a modal conflict if $L_1  \& \ldots \& \L_n$
and $L'_1 \& \ldots \L'_m$ can be true simultaneously for any system
state $G$.
\end{definition}

A modal authorization violation may be modeled with rules of
following form:

 {\ttfamily error $\leftarrow$
(($P_l,DS$)$\Rightarrow$ do($s$, $o$, $+a$)) $\&$
(($P_h,DS$)$\Rightarrow$ do($s$, $o$, $-a$)) }

\begin{definition}(Obligation Violations)\\
An obligation violation occurs, when a subject either does not
perform his or her obligations or does not perform obligations
correctly.\\
 \noindent Let $mustdo(s,a,q) \leftarrow L_1 \& \ldots
\& L_n$ be an obligation decision rule, where $s$ is a subject, $a$
is an action, $q$ is a post condition, and $L_1 \& \ldots \& L_n$ is
a precondition. Let $e_a$ be an effect of action $a$ asserted by the
ontology. When $L_1 \& \ldots \& L_n$ holds but $e_a ~\&~ q$ is not
satisfied, an obligation violation is indicated.
\end{definition}

We assume that prior to the time of compliance checking the subject
had sufficient time to perform obligations satisfactorily. Detection
of obligation violation may be modeled with rule of following form:

{\ttfamily error $\leftarrow$ $L_1 \& \ldots \& L_n ~\&~ \neg e_a
~\&~ \neg q$}

\begin{definition}(Resource Capability Conflict)\\
Resource capability conflict occurs when resources required to
perform an obligation does not exist.
\end{definition}

Let $DS=(O,I)$ be a data system, where $O$ is an ontology and $I$ is
set of objects in the system. A resource capability conflict may be
modeled with rules of following form:

{\ttfamily error $\leftarrow$ mustdo(s,a,q) $\&$ resource(a,r) $\&$
(r $\not \in I$)}

\begin{definition}(Modal Capability Conflict)\\
Modal capability conflict occurs when an obligation requires access
to certain resources, and the subject does not have the requisite
permissions.
\end{definition}

{\ttfamily error $\leftarrow$ mustdo(s,a,q) \&
$\neg$do(a,s,+execute)}

\begin{theorem}
Obligation and Authorization specification is a locally stratified
logic program, thus preserves the desirable properties given in
Theorem 1 of~\cite{Jajodia01}.
\end{theorem}
\begin{myproofsketch} Authorization specification language has been
extended by introducing new predicates. Table~\ref{table:strata}
shows that all atoms in the specification can be assigned a rank
such that no atom depends on an atom of greater rank or depends
negatively on one on equal or greater rank in any instantiated rule.
Proof of this theorem is similar to the proof of Theorem 1
of~\cite{Jajodia01}.
\end{myproofsketch}

We assume that the high-level policy does not contain positive
authorization rules. Any authorizations derived from $P_h$ must be
derived from obligation and derivation rules. If $D_h$ $\not
\subseteq$ $D_l$ then low-level security policy is prohibiting some
users from performing their obligations. This is a case of modal
capability conflict and the algorithm correctly returns false.

Obligations derived from the high-level security policy must occur
in a compliant low-level policy or the obligations must have been
satisfied. If the obligation is satisfied, the obligation
postcondition must be true and the effect of obligation action must
also be true. The compliance checking algorithm returns false, when
both the above conditions are not satisfied.

\begin{theorem}
Compliance checking algorithm (Alg.~\ref{alg:compliance}) terminates
and the algorithm returns false if low-level security policy and
system state is not compliant with the high-level security policy.
\end{theorem}
\begin{myproofsketch} The compliance checking algorithm computes decision
view of the high-level and low-level policy by evaluating their
obligation and authorization specifications, which are locally
stratified logic programs. The herbrand base of the obligation and
authorization specification is finite. Also, the variables used in
the rule head are bounded by the variables in the body of the rule.
The policy refinement process performs substitution of rules in the 
high-level policy until actions can not be further refined. Action refinements
in our framework cannot contain loops as the refinement are always
more specific. Therefore the number of times action refinement may
be performed is finite. We consider a finite DS, thus only a 
finite number of instantiations may occur; 
therefore Alg.~\ref{alg:compliance} terminates.

If the low-level security policy violates the high-level security
policy, the algorithm detects the violation and returns false. This
is proved by contradiction. Let us assume that the low-level
security policy $P_l$ violates high-level security policy $P_h$ and
the compliance checking algorithm returns true. The algorithm can
return true only if 1) $D_h \subseteq D_l$, and 2) for every
mustdo(s,a,q) predicate in $M_h$, either mustdo(s,a,q) is in $M_l$
or ($\sigma \Rightarrow q$ and $\sigma \Rightarrow e_a$). The
decision views ($P_l,DS$) and $(P_h,DS)_{ref}$ contain only
ground mustdo and do predicates. If ($\sigma \Rightarrow q$ and
$\sigma \Rightarrow e_a$), the obligation $a$ has already been
satisfied in $P_l$. As discussed the remainder of mustdo and do
predicates also occur in $(P_h,DS)_{ref}$. Hence, the low-level
security policy is compliant to high-level security policy. This is
in contradiction to initial assumption.
\end{myproofsketch}


\section{Conclusions}
\label{sec:conclude} In this paper we proposed a framework and
techniques to evaluate whether a low-level, implemented security
policy is compliant to a high-level policy.  Our method uses
organizational and security meta-data and a set of well-defined
operations to generate valid refinements of a given high-level
policy.  The implemented policy is compared to these refinements to
verify whether it is compliant to the high-level policy.  The
correctness of the compliance is based on the properties of the
refinement, that is the well-formedness of the refinement operators
and the validity of the compositions.

Although the basic concept presented in this work have been proposed
and used in other fields of research and development, e.g., software
engineering and programming languages, their relevance for
information security have not yet been fully evaluated.  Our aim is
to build upon these technologies to establish formal properties of
security policies.  This work constitutes our initial efforts on
incorporating results from software
refinement~\cite{Dijkstra71,Wirth71}, requirement analysis, and
process
algebra~\cite{Baeten90,Bergstra84,Bergstra85,Hoare78,Hoare85,Milner82,Glabbeek01}
in security policy verification. Our ongoing work includes analysis
of more complex policy refinements, usage of extensive
organizational meta-data, and bottom-up compliance verification. Our
goal is to develop methods and tools that will aid and simplify the
human evaluation process for compliance checking.

\section{Acknowledgements}
\noindent This work was partially supported by National Science
Foundation grant number IIS-0237782 and an IBM Summer Internship.
We would like to express our gratitude and thanks to Duminda
Wijesekera for his valuable comments and suggestions on policy
languages, refinement, and obligations.

\bibliographystyle{IEEEtran}
\bibliography{IEEEabrv,master}

\end{document}